# Couette-Poiseuille flow with partial slip and uniform cross flow for power-law fluids

Tarek M.A. El-Mistikawy
Department of Engineering Mathematics and Physics, Faculty of Engineering, Cairo University, Giza 12211, Egypt

## Abstract

Exact solutions are obtained for the steady flow of a power-law fluid between parallel plates with partial slip conditions and uniform cross flow. The problem is properly formulated and similarities are exploited. The exact solutions are obtained in terms of integrals which can be performed, in closed form, in special cases of the power-law index $n$. Solutions to cases of $n$ =½, 1, and 2; representing a pseudo-plastic, a Newtonian, and a dilatant fluid, respectively, are presented. Tendencies to corresponding degenerate cases in the literature are demonstrated. Depending on the strength of the cross flow and the pressure gradient, the flow may be of Couette type with convex, linear, or concave velocity profile; or of Poiseuille type. Borderline cases are identified. Moreover, a case in which the power-law model for the dilatant fluid fails is detected.

## 1. Introduction

Steady Couette and Poiseuille flows are classical problems of fluid mechanics. Both flows are considered here to take place between two parallel plates. Couette flow is incited by the motion of one of the plates with uniform speed. Poiseuille flow is incited by a constant streamwise pressure gradient. When both inciting agents are present we have the generalized Couette/Poiseuille (GCP) flow. Depending on the strengths of these agents, the flow can be of Couette type or Poiseuille type. The velocity profile of a Couette-type flow slopes monotonically, while that of a Poiseuille-type flow has a local extremum lying between the plates.

When the fluid is Newtonian, Couette and Poiseuille flows admit simple exact solutions of the governing Navier-stokes equations and no-slip boundary conditions [1]. The solution for the GCP flow is a linear combination of the solutions for the component flows. The linearity of the problems invited researchers to add new features.

Unsteadiness of the GCP flow was treated by Gopalan [2], who also allowed for uniform cross flow through porous plates and studied the associated steady heat transfer problem. His work encompasses the work of Fang [3] on Couette flow.

The cross flow has a significant impact on the flow. It destroys the linear velocity profile of Couette flow and the symmetric profile of Poiseuille flow. For the GCP flow, it can change a Couette-type flow to a Poiseuille-type flow and vice verse.

Allowing for slip conditions is another added feature, as the abovementioned works invoked the no-slip boundary conditions at the plates. Linear as well as nonlinear slip conditions were adopted [4,5,6]. Slip conditions are important in flows through micro-channels, flows of rarefied gasses, flows of slippery fluids such as slurries, polymers and blood.

Non-Newtonian fluids were also treated. Although the problems were then nonlinear, exact solutions were still possible to obtain- in the absence of cross flow- in case of power law fluids with no-slip conditions [6,7,8] and with slip conditions [6,9,10], and in case of second grade fluids [6,11 12].

In the present article, we derive exact solutions, in the form of integrals, for the steady GCP flow of power law fluids allowing for cross flow and linear slip conditions. The integrals are performed for special cases of pseudo plastic, Newtonian and dilatant fluids. Illustrative results are presented, when fixing the wall speed while varying the cross flow speed and/or the pressure gradient.

Recently, the author has become aware of Ref. [13], which handled, in almost the same way, the same problem with the no-slip conditions, when fixing the cross flow speed and the pressure gradient while varying the wall speed. In Appendix C, we extend the work of Ref. [13] allowing for partial slip. However, the current symbols, terminology and approach are in use, instead of those of Ref. [13].

## 2. Problem formulation

The steady Couette-Poiseuille flow of a power-law fluid with partial slip conditions and uniform cross flow is governed by the following problem, in dimensional variables marked with primes.

$$\frac{d\tau'}{dy'} = \rho' V' \frac{du'}{dy'} + \frac{dp'}{dx'}$$

$$u' - \sigma_0' \tau' = w' \qquad \text{at } y' = 0$$

$$u' + \sigma_1' \tau' = 0 \qquad \text{at } y' = h'$$

where $h'$ is the distance between the two walls, $y'$ measures distances from the moving wall toward the fixed wall, $x'$ measures distances parallel to the walls in the direction of the wall speed $w'$ and flow velocity $u'(y')$, $p'$ is the pressure, and $\tau'$ is the shear stress given by the power law

$$\tau' = \eta' \left|\frac{du'}{dy'}\right|^{n-1} \frac{du'}{dy'}$$

and of interest is the flow rate

$$q' = \int_0^{h'} \rho' u' dy'$$

Assumed constant are the wall speed $w'$, the cross flow velocity $V'$ in the $y'$ direction, and the pressure gradient $dp'/dx'$; and for a given fluid, the density $\rho'$, the viscosity parameter $\eta'$, the power-law index $n$, and the slip coefficients $\sigma_0'$ and $\sigma_1'$.

Introduce the new non-dimensional variables

$y = y'/h'$, $u = u'/U'$, $\tau = \tau'/T'$, and $q = q'/(\rho' U' h')$;

and define the new parameters

$w = w'/U'$, $V = \rho' U' V'/T'$, $\pi = -(dp'/dx')h'/T'$, and $(\sigma_0, \sigma_1) = (\sigma_0', \sigma_1')T'/U'$;

where

$$T' = \eta' (U'/h')^n$$

with a suitably chosen characteristic speed

$U' = w'$, $|V'|$, or $(|dp'/dx'| h'/\rho')^{1/2}$.

(Reference [13] uses $U' = \frac{n}{n+1}(-\frac{1}{\eta'}\frac{dp'}{dx'})^{\frac{1}{n}}h'^{\frac{n+1}{n}}$ so that $\pi = (1+\frac{1}{n})^n > 0$, enforcing a favorable pressure gradient $\frac{dp'}{dx'} < 0$ along the $x'-$ axis.)

The problem takes the following form, where the subscripts 0 and 1 refer to the moving and fixed walls, respectively.

$$\frac{d\tau}{dy} = V\frac{du}{dy} - \pi \quad (1)$$

$$u_0 - \sigma_0\tau_0 = w \quad (2)$$

$$u_1 + \sigma_1\tau_1 = 0 \quad (3)$$

$$\tau = \left|\frac{du}{dy}\right|^{n-1}\frac{du}{dy} \quad (4)$$

with

$$q = \int_0^1 u dy \quad (5)$$

The problem is unaltered under the transformation

$$[\pi, V, \sigma_0, \sigma_1, u(y), \tau(y), q] \rightarrow [-\pi, -V, \sigma_1, \sigma_0, w - u(1-y), \tau(1-y), w - q].$$

Therefore, only cases of $\pi \geq 0$ need to be considered. In these cases, the flow cannot be of backward Poiseuille type regardless of the value of $V$ {Appendix B, Note #1}.

The analysis presented below is for $\pi > 0$ and $V \neq 0$. The results for the cases of $\pi = 0$ and/or $V = 0$ are given in Appendix A.

## 3. Exact solution

Equation (1) integrates to

$$\tau = \tau_i + V(u - u_i) - \pi(y - i), \quad i = 0,1 \quad (6)$$

so that, with $y = 1$ and $i = 0$

$$\tau_1 = \tau_0 + V(u_1 - u_0) - \pi \quad (7)$$

Substituting for $u_0$ and $u_1$ from Conditions (2) and (3), respectively, and rearranging give

$$(1 - V\sigma_0)\tau_0 - (1 + V\sigma_1)\tau_1 = Vw + \pi \quad (8)$$

To proceed further, we need to consider cases of positive and negative $du/dy$ separately.

### 3.1. Case of positive $du/dy$

For $du/dy > 0$, Eq. (4) gives $\tau = (du/dy)^n > 0$ so that $du/dy = \tau^{1/n}$. Then, Eq. (1) becomes

$$\frac{d\tau}{dy} = V\tau^{1/n} - \pi$$

which integrates to give $y$ in terms of $\tau$ as

$$y = i + \int_{\tau_i}^{\tau} \frac{d\hat{\tau}}{V\hat{\tau}^{1/n} - \pi}, \qquad i = 0,1 \tag{9$_+$}$$

### 3.2. Case of negative $du/dy$

For $du/dy < 0$, Eq. (4) gives $\hat{\tau} = -\tau = (-du/dy)^n > 0$ so that $\frac{du}{dy} = -\hat{\tau}^{1/n}$. Then, Eq. (1) becomes

$$\frac{d\hat{\tau}}{dy} = V\hat{\tau}^{1/n} + \pi$$

which integrates to give $y$ in terms of $\tau$ as

$$y = i + \int_{-\tau_i}^{-\tau} \frac{d\hat{\tau}}{V\hat{\tau}^{1/n} + \pi}, \qquad i = 0,1 \tag{9$_-$}$$

## 4. Couette-type flows

Couette-type flows are characterized by $du/dy < 0$ for $0 < y < 1$ {Appendix B, Note #2}, so that Eq. (9$_-$) applies, all through; leading, in particular, to

$$\int_{-\tau_0}^{-\tau_1} \frac{d\hat{\tau}}{V\hat{\tau}^{1/n} + \pi} = 1 \tag{10}$$

Now, given a set of flow parameters $w$, $V$, $\pi$, $\sigma_0$, and $\sigma_1$, we solve Eqs. (8) and (10) for $\tau_0$ and$\tau_1$. Then, $u_0$ and $u_1$ are obtained from Eqs. (2) and (3), respectively. Next, with $i = 0$ or 1, Eq. (9$_-$) is solved for $\tau$, then Eq. (6) is solved for $u$; to get each, in terms of $y$. Finally, $q$ is determined from Eq. (5).

When $V \geq 0$ the velocity profile is convex with its slope monotonically decreasing {Appendix B, Note #3}, with increasing $y$. As $V$ gradually decreases below zero, the profile changes from being convex to being linear then concave.

The value $V = V_L < 0$, at which the profile is linear, can be shown to satisfy the following equation {Appendix B, Note #4}

$$(\sigma_0 + \sigma_1)(-\frac{\pi}{V_L})^n + (-\frac{\pi}{V_L}) - w = 0 \tag{11}$$

The flow seizes to be of Couette type when $V$ is above a finite value $V_F$ at which $\tau_0 = 0$; becoming of forward Poiseuille type with $\tau_0 > 0$.

The integral, appearing with different limits in Eqs. (9$_-$) and (10), can be evaluated for some values of $n$; three of which are considered below.

### 4.1. The case of a shear thinning (pseudo-plastic) fluid ($n$= ½)

For $V > 0$, Eq. (10) gives

$$\tan^{-1}\frac{-\tau_1}{\sqrt{\pi/V}} - \tan^{-1}\frac{-\tau_0}{\sqrt{\pi/V}} = \sqrt{\pi V} \tag{12}$$

On the other hand, Eq. (9.) with $i = 1$ gives

$$y = 1 + \frac{1}{\sqrt{\pi V}}[\tan^{-1}\frac{-\tau}{\sqrt{\pi/V}} - \tan^{-1}\frac{-\tau_1}{\sqrt{\pi/V}}]$$

which can be rearranged to give

$$\tau = -\sqrt{\pi/V}\tan[\tan^{-1}\frac{-\tau_1}{\sqrt{\pi/V}} - \sqrt{\pi V}(1-y)] \tag{13}$$

Setting $\tau_0 = 0$ in Eqs. (8) and (12) then eliminating $\tau_1$, it is found that $V_F > 0$ satisfies

$$\sqrt{\pi/V_F}\tan\sqrt{\pi V_F} - \frac{V_F w + \pi}{1 + V_F\sigma_1} = 0 \tag{14}$$

For $V < 0$, Eq. (10) gives the following equation

$$\ln\frac{\sqrt{-\pi/V} + (-\tau_1)}{\sqrt{-\pi/V} + (-\tau_0)} - \ln\frac{\sqrt{-\pi/V} - (-\tau_1)}{\sqrt{-\pi/V} - (-\tau_0)} = 2\sqrt{-\pi V} \tag{15}$$

Similarly, Eq. (9.) with $i = 0$ gives

$$\ln\frac{\sqrt{-\pi/V} + (-\tau)}{\sqrt{-\pi/V} + (-\tau_0)} - \ln\frac{\sqrt{-\pi/V} - (-\tau)}{\sqrt{-\pi/V} - (-\tau_0)} = 2\sqrt{-\pi V}y$$

which can be manipulated to give

$$\tau = \sqrt{-\pi/V}\frac{(\frac{\tau_0}{\sqrt{-\pi/V}}) - \tanh\sqrt{-\pi V}\mathrm{y}}{1 - (\frac{\tau_0}{\sqrt{-\pi/V}})\tanh\sqrt{-\pi V}\mathrm{y}} \tag{16}$$

Setting $\tau_0 = 0$ in Eqs. (8) and (15) then eliminating $\tau_1$, it is found that $V_F$ satisfies

$$\sqrt{-\pi/V_F}\tanh\sqrt{-\pi V_F} - \frac{V_F w + \pi}{1 + V_F\sigma_1} = 0 \tag{17}$$

while Eq. (11) gives $V_L$ as

$$V_L = -\pi/[-\tfrac{1}{2}(\sigma_0 + \sigma_1) + \sqrt{\tfrac{1}{4}(\sigma_0 + \sigma_1)^2 + w}]^2 \tag{18}$$

### 4.2. The case of a Newtonian fluid (*n*=1)

Eqs. (1)-(3), with $\tau = du/dy$ are solved to give

$$\tau = \frac{\pi}{V} - \frac{\pi(\sigma_0 + \sigma_1) + (Vw + \pi)}{(1 + V\sigma_1)e^V - (1 - V\sigma_0)} e^{Vy} \tag{19}$$

from which $\tau_0$ and $\tau_1$ and consequently $u_0$ and $u_1$ can be readily obtained.

The value of $V_F$ satisfies

$$\pi(e^{V_F} - 1)(1 + V_F\sigma_1) - V_F(V_F w + \pi) = 0 \tag{20}$$

while Eq. (11) gives $V_L$ as

$$V_L = -(1 + \sigma_0 + \sigma_1)\pi/w \tag{21}$$

### 4.3. The case of a shear thickening (dilatant) fluid ($n$=2)

Equation (10) gives the following equation

$$\sqrt{-\tau_1} - \sqrt{-\tau_0} - \frac{\pi}{V}\ln\frac{\sqrt{-\tau_1} + \pi/V}{\sqrt{-\tau_0} + \pi/V} = \frac{V}{2} \tag{22}$$

which, with the help of Eq. (8), can be solved for $\tau_0$ and $\tau_1$.

On the other hand, Eq. (9.) gives the following equation relating $\tau$ to $y$.

$$y = \frac{2}{V}[\sqrt{-\tau} - \sqrt{-\tau_0} - \frac{\pi}{V}\ln\frac{\sqrt{-\tau} + \pi/V}{\sqrt{-\tau_0} + \pi/V}] \tag{23}$$

The value of $V_F$ satisfies

$$\sqrt{\frac{V_F w + \pi}{1 + V_F\sigma_1}} - \frac{\pi}{V_F}\ln[1 + \frac{V_F}{\pi}\sqrt{\frac{V_F w + \pi}{1 + V_F\sigma_1}}] - \frac{V_F}{2} = 0 \tag{24}$$

while Eq. (11) gives $V_L$ as

$$V_L = -\frac{\pi}{2w}\left[\sqrt{1 + 4(\sigma_0 + \sigma_1)w} + 1\right] = \frac{-2\pi(\sigma_0 + \sigma_1)}{\sqrt{1 + 4(\sigma_0 + \sigma_1)w} - 1} \tag{25}$$

In all three cases of $n$ =½, 1 and 2, Eq. (6) gives

$$u = w + \sigma_0\tau_0 + \frac{1}{V}[\tau - \tau_0 + \pi y] = -\sigma_1\tau_1 + \frac{1}{V}[\tau - \tau_1 + \pi(y - 1)] \tag{26}$$

where Conditions (2) and (3) have been used.

The integral in Eq. (5) can be evaluated leading to

$$n = \tfrac{1}{2}\colon q = \hat{q} - \frac{1}{2V^2}\ln\frac{{\tau_1}^2 + \pi/V}{{\tau_0}^2 + \pi/V} \tag{27a}$$

$$n = 1: q = \hat{q} + \frac{1}{V^2}[(\tau_1 - \tau_0) + \frac{\pi}{V}\ln\frac{\tau_1 - \pi/V}{\tau_0 - \pi/V}] \tag{27b}$$

$$n = 2: q = \hat{q} + \frac{2}{V^2}[(\frac{\pi}{V})^3 \sum_{k=1}^{3} \frac{(-\tau_1)^{k/2} - (-\tau_0)^{k/2}}{k(-\pi/V)^k} + \frac{\pi}{V}\ln\frac{\sqrt{-\tau_1} + \pi/V}{\sqrt{-\tau_0} + \pi/V}] \tag{27c}$$

where

$$\hat{q} = w + \sigma_0\tau_0 + \frac{\pi - 2\tau_0}{2V} = -\sigma_1\tau_1 - \frac{\pi + 2\tau_1}{2V} \tag{27d}$$

## 5. Forward Poiseuille-type flows

Poiseuille-type flows, for $\pi > 0$ and $V > V_F$, are characterized by $\tau_0 > 0$ and $\tau_1 < 0$. The velocity profile has two branches, which join at $y = y^*$, where $u = u^*$ and $du/dy = 0$.

For $0 \leq y \leq y^*$, $du/dy \geq 0$, then Eq. ($9_+$) gives

$$y = \int_{\tau_0}^{\tau} \frac{d\hat{\tau}}{V\hat{\tau}^{1/n} - \pi} \tag{28$_+$}$$

so that

$$y^* = \int_{\tau_0}^{0} \frac{d\hat{\tau}}{V\hat{\tau}^{1/n} - \pi} \tag{29$_+$}$$

For $y^* \leq y \leq 1$, $du/dy \leq 0$, then Eq. ($9_-$) gives

$$y = 1 + \int_{-\tau_1}^{-\tau} \frac{d\hat{\tau}}{V\hat{\tau}^{1/n} + \pi} \tag{28$_-$}$$

so that

$$y^* = 1 + \int_{-\tau_1}^{0} \frac{d\hat{\tau}}{V\hat{\tau}^{1/n} + \pi} \tag{29$_-$}$$

Equating the right-hand-sides of Eqs. ($29_+$) and ($29_-$) leads to

$$\int_{\tau_0}^{0} \frac{d\hat{\tau}}{V\hat{\tau}^{1/n} - \pi} = 1 + \int_{-\tau_1}^{0} \frac{d\hat{\tau}}{V\hat{\tau}^{1/n} + \pi} \tag{30}$$

Now, given a set of flow parameters $w$, $V$, $\pi$, $\sigma_0$, and $\sigma_1$, we solve Eqs. (8) and (30) for $\tau_0$ and $\tau_1$. Then, $u_0$, and $u_1$ are obtained from Eqs. (2) and (3), respectively. $y^*$ is determined from Eq. ($29_+$) or ($29_-$).

For $0 \leq y \leq y^*$, Eq. ($28_+$) gives $\tau$ in terms of $y$, while Eq. (6) with $i = 0$ gives $u$ in terms of $y$ as

$$u = u_0 + \frac{1}{V}(\tau - \tau_0 + \pi y) \tag{31$_+$}$$

For $y^* \le y \le 1$, Eq. (28$_-$) gives $\tau$ in terms of $y$, while Eq. (6) with $i = 1$ gives $u$ in terms of $y$ as

$$u = u_1 + \frac{1}{V}[\tau - \tau_1 + \pi(y - 1)] \tag{31$_-$}$$

$u^*$ is determined from Eq. (31$_+$) or (31$_-$) with $y = y^*$ and $\tau = 0$.

The three cases of $n$ =½, 1, 2 are considered below.

### 5.1. The case of a pseudo-plastic fluid ($n$ =½)

For $V > 0$, Eq. (30) gives

$$\tan^{-1}\frac{-\tau_1}{\sqrt{\pi/V}} + \tanh^{-1}\frac{\tau_0}{\sqrt{\pi/V}} = \sqrt{\pi V} \tag{32}$$

which, with the help of Eq. (8), can be solved for $\tau_0 > 0$ and $\tau_1 < 0$.

For $0 \le y \le y^*$, Eq. (28$_+$) gives $\tau$ in terms of $y$ as

$$\tau = \sqrt{\pi/V}\tanh[\tanh^{-1}\frac{\tau_0}{\sqrt{\pi/V}} - \sqrt{\pi V}y] \tag{33$_+$}$$

For $y^* \le y \le 1$, Eq. (28$_-$) gives $\tau$ in terms of $y$ as

$$\tau = \sqrt{\pi/V}\tan[\tan^{-1}\frac{\tau_1}{\sqrt{\pi/V}} + \sqrt{\pi V}(1 - y)] \tag{33$_-$}$$

where $y^*$ is given by

$$y^* = \frac{1}{\sqrt{\pi V}}\tanh^{-1}\frac{\tau_0}{\sqrt{\pi/V}} = 1 - \frac{1}{\sqrt{\pi V}}\tan^{-1}\frac{-\tau_1}{\sqrt{\pi/V}} \tag{34}$$

For $V < 0$, Eq. (30) gives

$$\tanh^{-1}\frac{-\tau_1}{\sqrt{-\pi/V}} + \tan^{-1}\frac{\tau_0}{\sqrt{-\pi/V}} = \sqrt{-\pi V} \tag{35}$$

which, with the help of Eq. (8), can be solved for $\tau_0 > 0$ and $\tau_1 < 0$.

For $0 \le y \le y^*$, Eq. (28$_+$) gives $\tau$ in terms of $y$ as

$$\tau = \sqrt{-\pi/V}\tan[\tan^{-1}\frac{\tau_0}{\sqrt{-\pi/V}} - \sqrt{-\pi V}y] \tag{36$_+$}$$

For $y^* \le y \le 1$, Eq. (28$_-$) gives $\tau$ in terms of $y$ as

$$\tau = \sqrt{-\pi/V}\tanh[\tanh^{-1}\frac{\tau_1}{\sqrt{-\pi/V}} - \sqrt{-\pi V}(1-y)] \tag{36-}$$

where $y^*$ is given by

$$y^* = \frac{1}{\sqrt{-\pi V}}\tan^{-1}\frac{\tau_0}{\sqrt{-\pi/V}} = 1 + \frac{1}{\sqrt{-\pi V}}\tanh^{-1}\frac{-\tau_1}{\sqrt{-\pi/V}} \tag{37}$$

### 5.2. The case of a Newtonian fluid ($n = 1$)

Equation (19) still gives $\tau$ in terms of $y$. Setting $\tau = 0$ in Eq. (19) results in

$$y^* = \frac{1}{V}\ln[\frac{\pi}{V}\frac{(1+V\sigma_1)e^V - (1-V\sigma_0)}{(1+\sigma_0+\sigma_1)\pi + Vw}] \tag{38}$$

### 5.3. The case of a dilatant fluid ($n = 2$)

Equation (30) gives

$$\sqrt{-\tau_1} - \sqrt{\tau_0} + \frac{\pi}{V}[\ln\frac{\pi/V}{\pi/V - \sqrt{\tau_0}} + \ln\frac{\pi/V}{\pi/V + \sqrt{-\tau_1}}] = \frac{V}{2} \tag{39}$$

which, with the help of Eq. (8), can be solved for $\tau_0 > 0$ and $\tau_1 < 0$.

For $0 \le y \le y^*$, Eq. (28+) relates $\tau$ to $y$ by,

$$y = \frac{2}{V}[\sqrt{\tau} - \sqrt{\tau_0} + \frac{\pi}{V}\ln\frac{\pi/V - \sqrt{\tau}}{\pi/V - \sqrt{\tau_0}}] \tag{40+}$$

For $y^* \le y \le 1$, Eq. (28-) relates $\tau$ to $y$ by

$$y = 1 + \frac{2}{V}[\sqrt{-\tau} - \sqrt{-\tau_1} - \frac{\pi}{V}\ln\frac{\pi/V + \sqrt{-\tau}}{\pi/V + \sqrt{-\tau_1}}] \tag{40-}$$

where $y^*$ is given by

$$y^* = \frac{2}{V}[-\sqrt{\tau_0} + \frac{\pi}{V}\ln\frac{\pi/V}{\pi/V - \sqrt{\tau_0}}] = 1 - \frac{2}{V}[\sqrt{-\tau_1} + \frac{\pi}{V}\ln\frac{\pi/V}{\pi/V + \sqrt{-\tau_1}}] \tag{41}$$

In all three cases of $n =$½, 1 and 2, $u$ is given by Eq. (26). For the flow rate, we have

$$n = \tfrac{1}{2}: q = \hat{q} + \frac{1}{2V^2}[\ln\frac{\pi/V}{\pi/V - {\tau_0}^2} + \ln\frac{\pi/V}{\pi/V + {\tau_1}^2}] \tag{42a}$$

$$n = 1: q = \hat{q} + \frac{1}{V^2}[(\tau_1 - \tau_0) + \frac{\pi}{V}\ln\frac{\pi/V - \tau_1}{\pi/V - \tau_0}] \tag{42b}$$

$$n = 2: q = \hat{q} + \frac{2}{V^2}(\frac{\pi}{V})^3[\sum_{k=1}^{3}\frac{(-1)^k(-\tau_1)^{k/2} - (\tau_0)^{k/2}}{k(\pi/V)^k} + \ln\frac{\pi/V + \sqrt{-\tau_1}}{\pi/V - \sqrt{-\tau_0}}] \tag{42c}$$

where $\hat{q}$ is given by Eq. (27d).

## 6. Sample Numerical Results

The analysis of Eqs. (1)-(5) presented above gives exact (explicit or implicit) expressions for some flow quantities of interest, in the case of Couette-Poiseuille flow of non-Newtonian fluids, which are modeled by the power law and subjected to slip conditions and cross flow. They can be used to generate numerical results, for specific values of the flow parameters ($n$, $\sigma_0$, $\sigma_1$ $w$, $V$, $\pi$), as may be needed.

The numerical results presented below are intended to demonstrate trends; revealing the behavior of the power-law fluids. Chosen are three values of the power-law index $n =$ ½, 1, and 2; representing a shear thinning, a Newtonian, and a shear-thickening fluids, respectively. Unless otherwise stated, all presented results are for $w = 1$ (i.e. $U' = w'$) and $\sigma = \sigma_0 = \sigma_1 = 0.1$.}

To start with, the borderline values (with subscript $F$), corresponding to $\tau_0 = 0$, which separates Couette-type flows from Poiseuille-type flows are given in Table 1. For different values of the pressure gradient parameter $\pi$, we give the cross flow velocity $V_F$, and for $V = 0$, we give $\pi_F$. Also given is the ratio $V_L/\pi$ corresponding to a linear Couette flow profile.

Table 1: Borderline Values

| $n$ | ½ | 1 | 2 |
|---|---|---|---|
| $\pi$ | - - - - - - - - - - | $V_F$ | - - - - - - - - - - |
| 0.0 | ∞ | ∞ | 3.062257748[a] |
| 0.2 | 11.973343243 | 4.184795653 | 2.424177310 |
| 0.4 | 5.639565075 | 3.054167033 | 1.979868025 |
| 0.6 | 3.416315901 | 2.317074116 | 1.592075135 |
| 0.8 | 2.219173150 | 1.745386650 | 1.236733646 |
| 1.0 | 1.431141547 | 1.265601652 | 0.903358068 |
| 2.0 | -0.693223997 | -0.529109075 | -0.578588592 |
| 3.0 | -2.052512053 | -1.925273414 | -1.899681713 |
| 4.0 | -3.252913766 | -3.176667773 | -3.140021852 |
| 5.0 | -4.403800804 | -4.361971314 | -4.332798338 |
| $\pi_F$ | 1.588533865 | 1.666666667 | 1.591001773 |
| $V_L/\pi$ | -1.220997512 | -1.200000000 | -1.170820393 |

a. See Eq. (A12) and the lines that follow, in Appendix A.

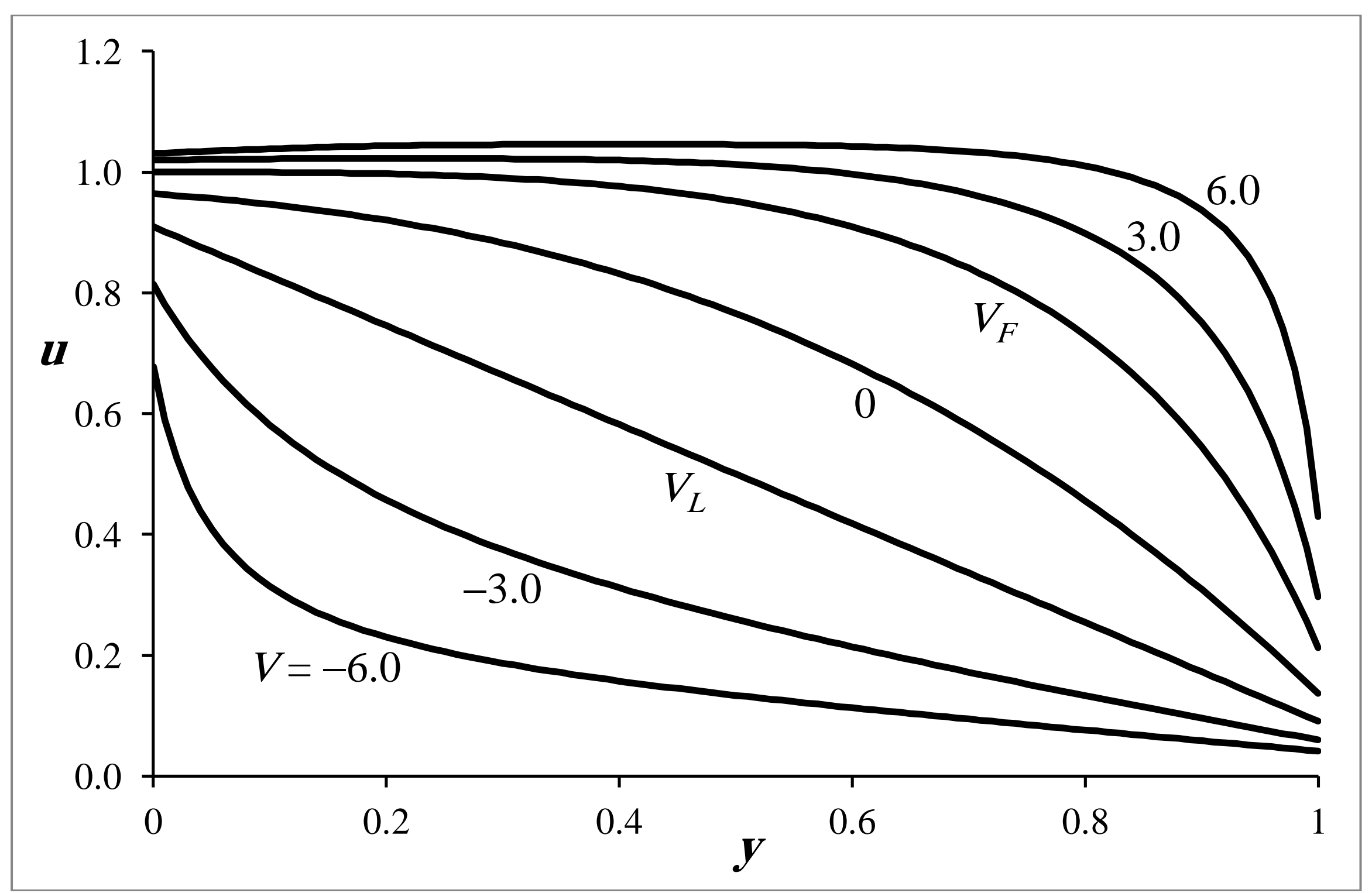

Fig. 1a: Velocity profiles for a shear-thinning fluid, $n$ = ½; $\pi$ =1.

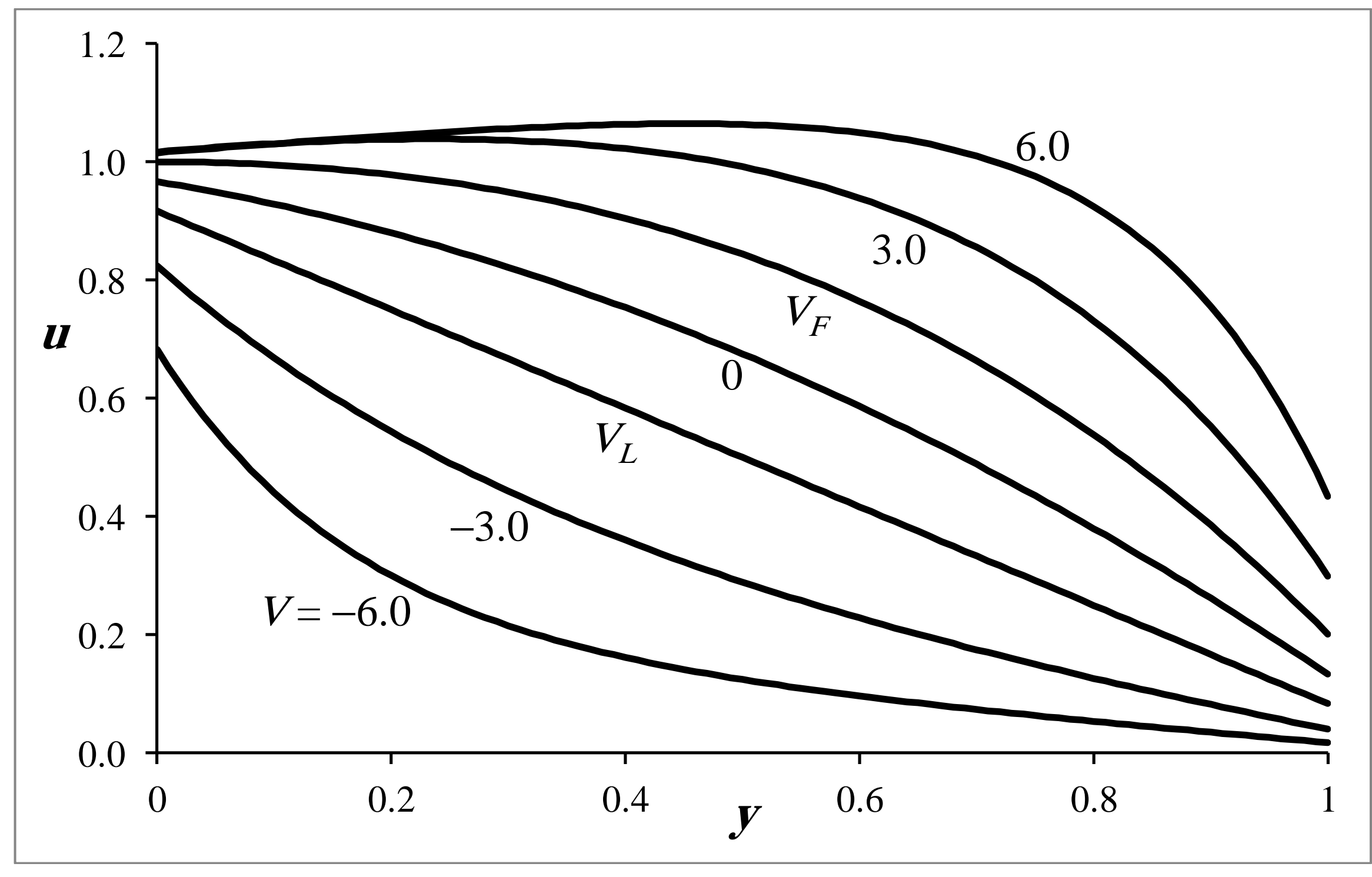

Fig. 1b: Velocity profiles for a Newtonian fluid, $n$ =1; $\pi$ =1.

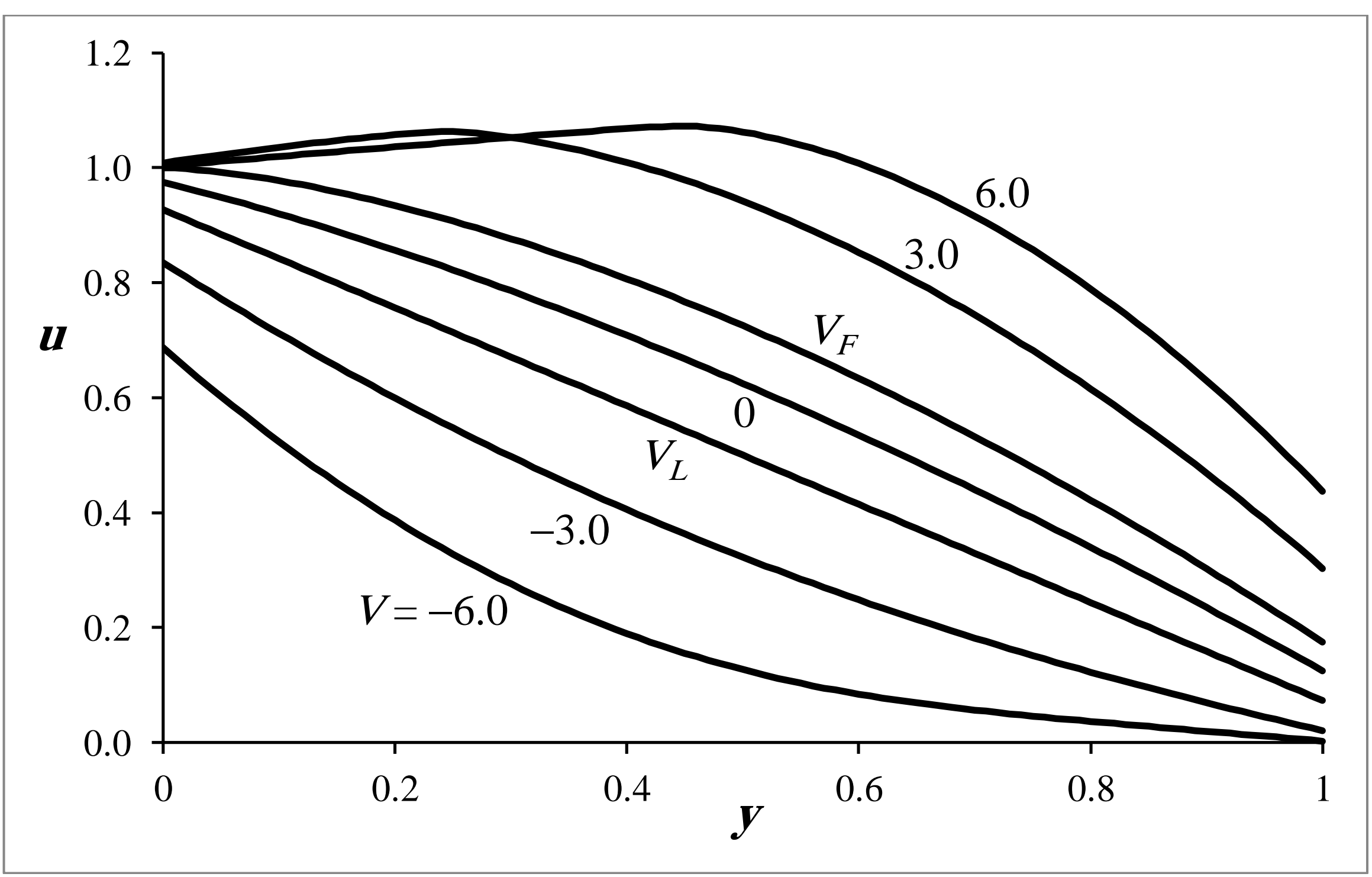

Fig. 1c: Velocity profiles for a shear-thickening fluid, $n$ =2; $\pi$ =1.

The velocity profiles for progressing values of $V$ that correspond to Couette-type, borderline, and Poiseuille-type flows; at the representative value $\pi$ =1; are shown in Figs. 1. The profiles exhibit qualitative differences- among cases of $n$ =½, 1, and 2- that are obvious at large values of $V > V_F$. The shear-thinning fluid ($n$ =½) develops a plateau-type profile ending with a steep drop toward the fixed wall, while the shear-thickening fluid ($n$ =2) develops a mountainous-type profile with clear crest, which moves toward the fixed wall as $V$ increases. The Newtonian profiles are in between.

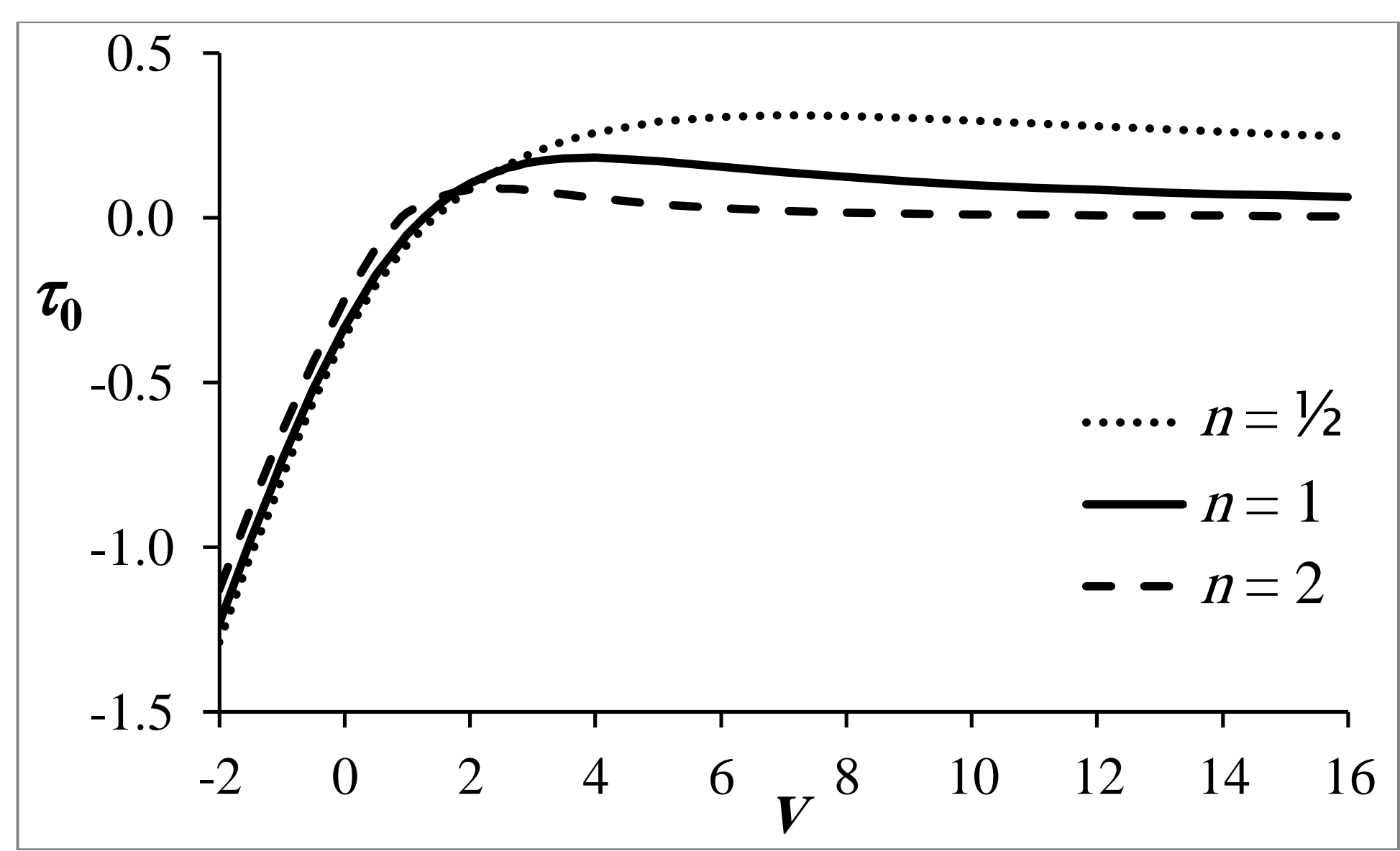

Fig. 2a: Variation of $\tau_0$ with $V$; $\pi$ =1.

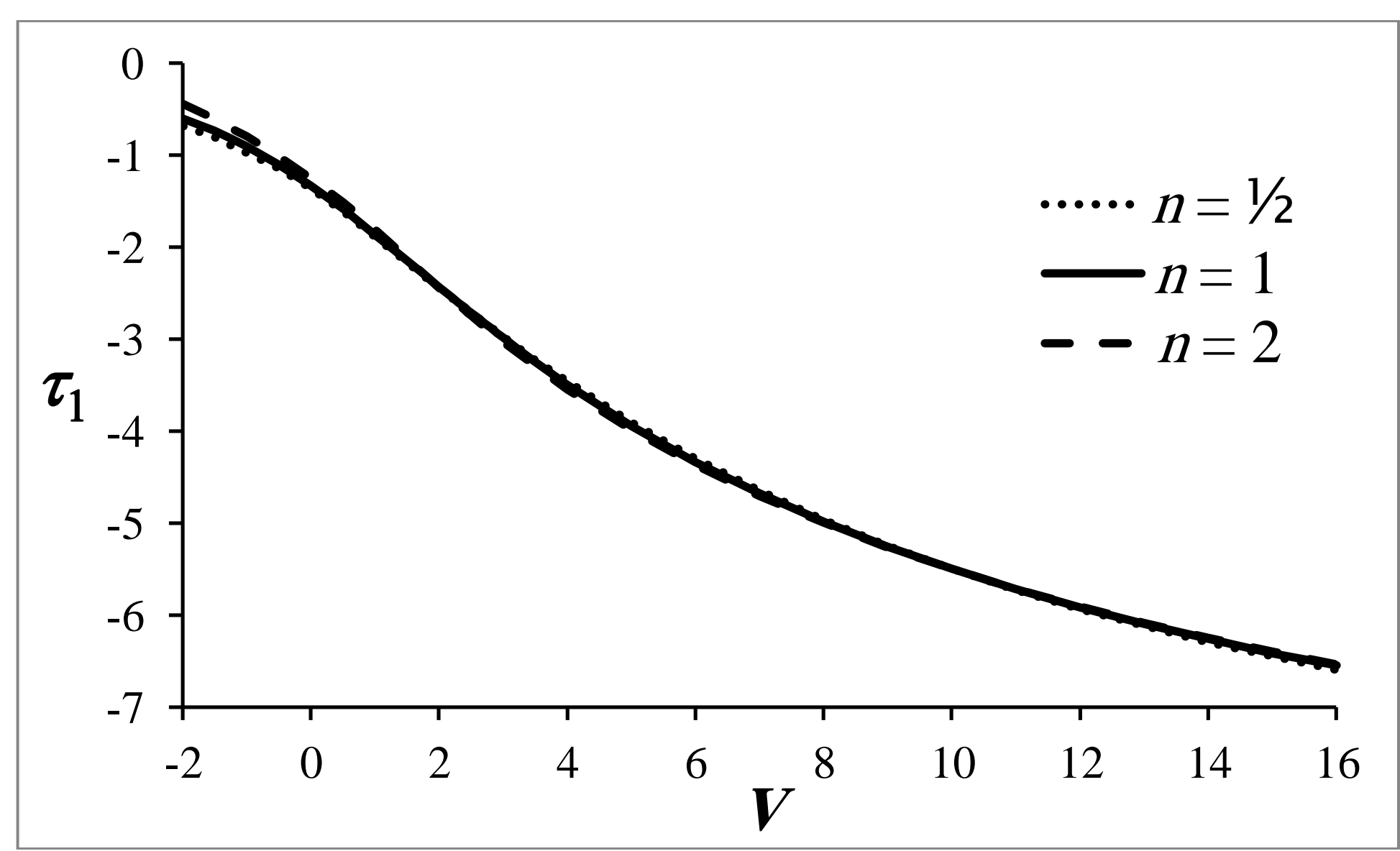

Fig. 2b: Variation of $\tau_1$ with $V$; $\pi$ =1.

The tendency of the fluids to slip is directly related to the shear stress at both walls {Conditions (2) and (3)}. It is measured by the slip velocities $u_0 - w = \sigma_0\tau_0$ up the moving wall and $u_1 = -\sigma_1\tau_1$ up the fixed wall. Along the moving wall, the fluids slip backwardly ($u_0 < w$) when the flow is of Couette type ($\tau_0 < 0$), and forwardly ($u_0 > w$) when the flow is of Poiseuille type ($\tau_0 > 0$). As $V$ increases, the speed of the backward slip decreases; vanishing when $V = V_F$, while the speed of the forward slip increases; reaching a maximum, then decreases. Along the fixed wall ($\tau_1 < 0$), the fluids slip forwardly ($u_1 > 0$). As $V$ increases, the speed of this forward slip decreases monotonically. The variations of $\tau_0$ and $\tau_1$ with $V$ are demonstrated in Figs. 2, in the case of $\pi$ =1.

Another quantity of interest is the flow rate $q$. Figs. 3 depict the variation of $q$ with $\pi$, when $V$ =0, while Figs. 4 depict the variation of $q$ with $V$, when $\pi$ =2; in both cases of Couette-type and Poiseuille-type flows.

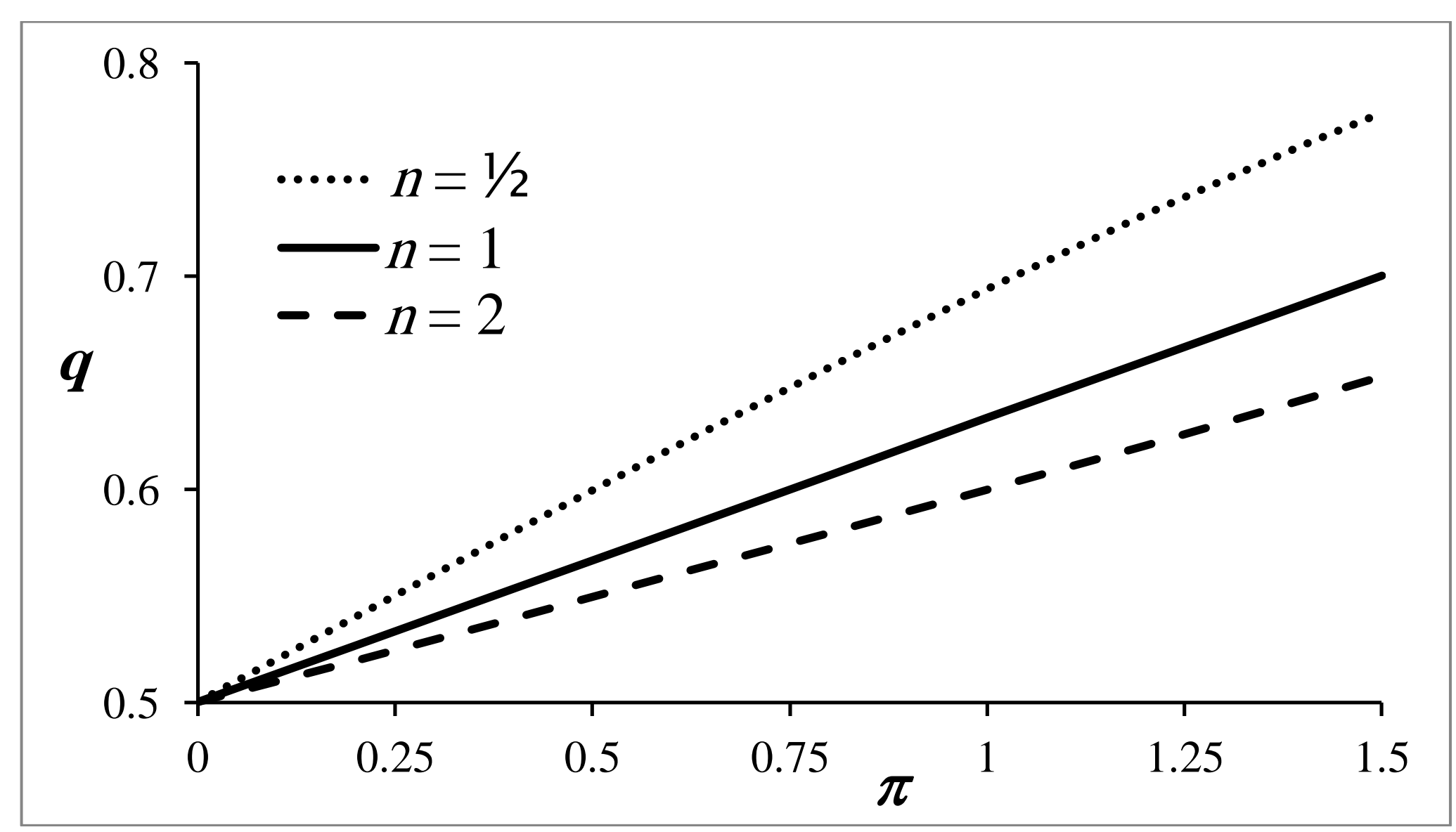

Fig. 3a: Flow rate vs. pressure gradient parameter, in Couette flow; $V = 0$.

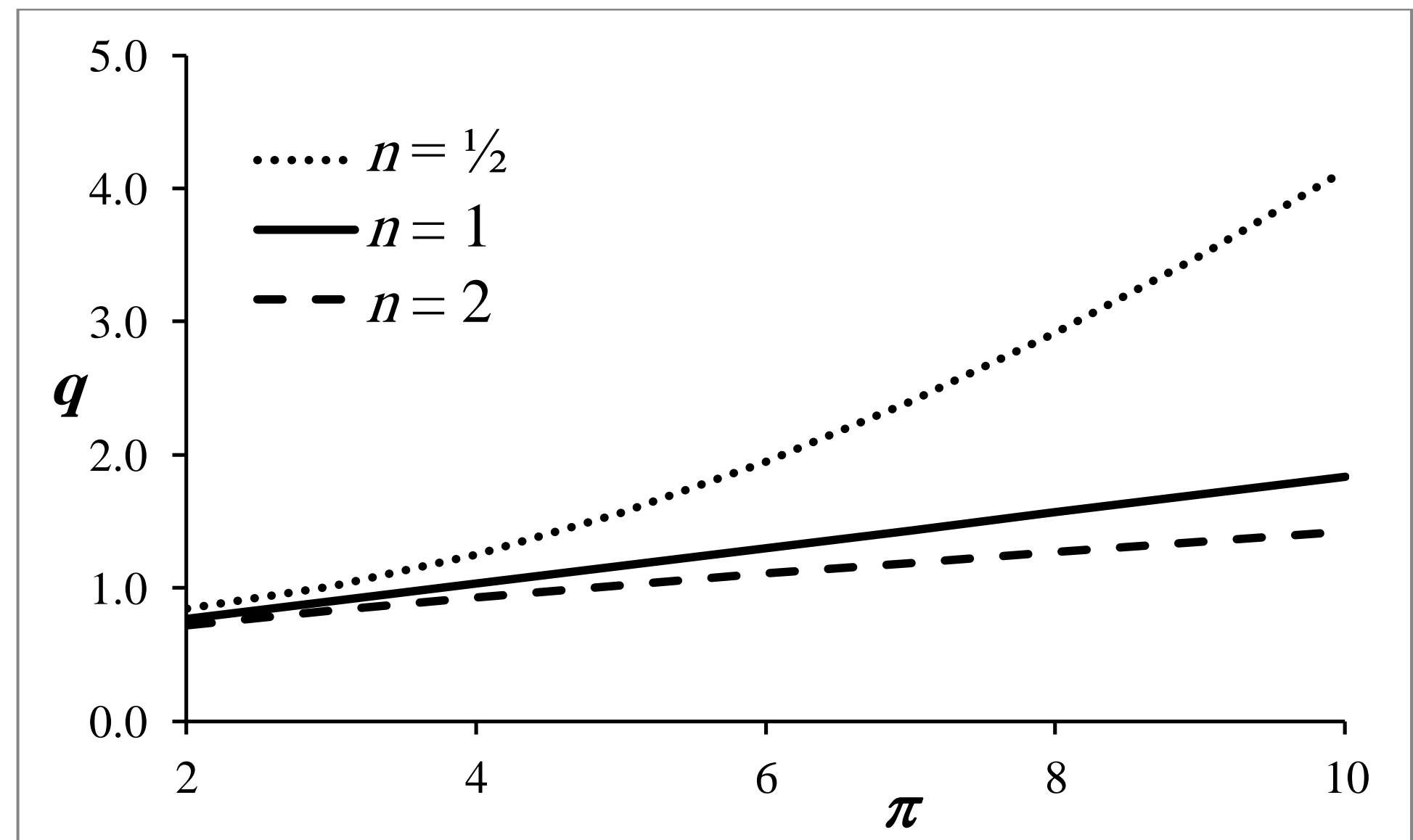

Fig. 3b: Flow rate vs. pressure gradient parameter, in Poiseuille flow; $V$ =0.

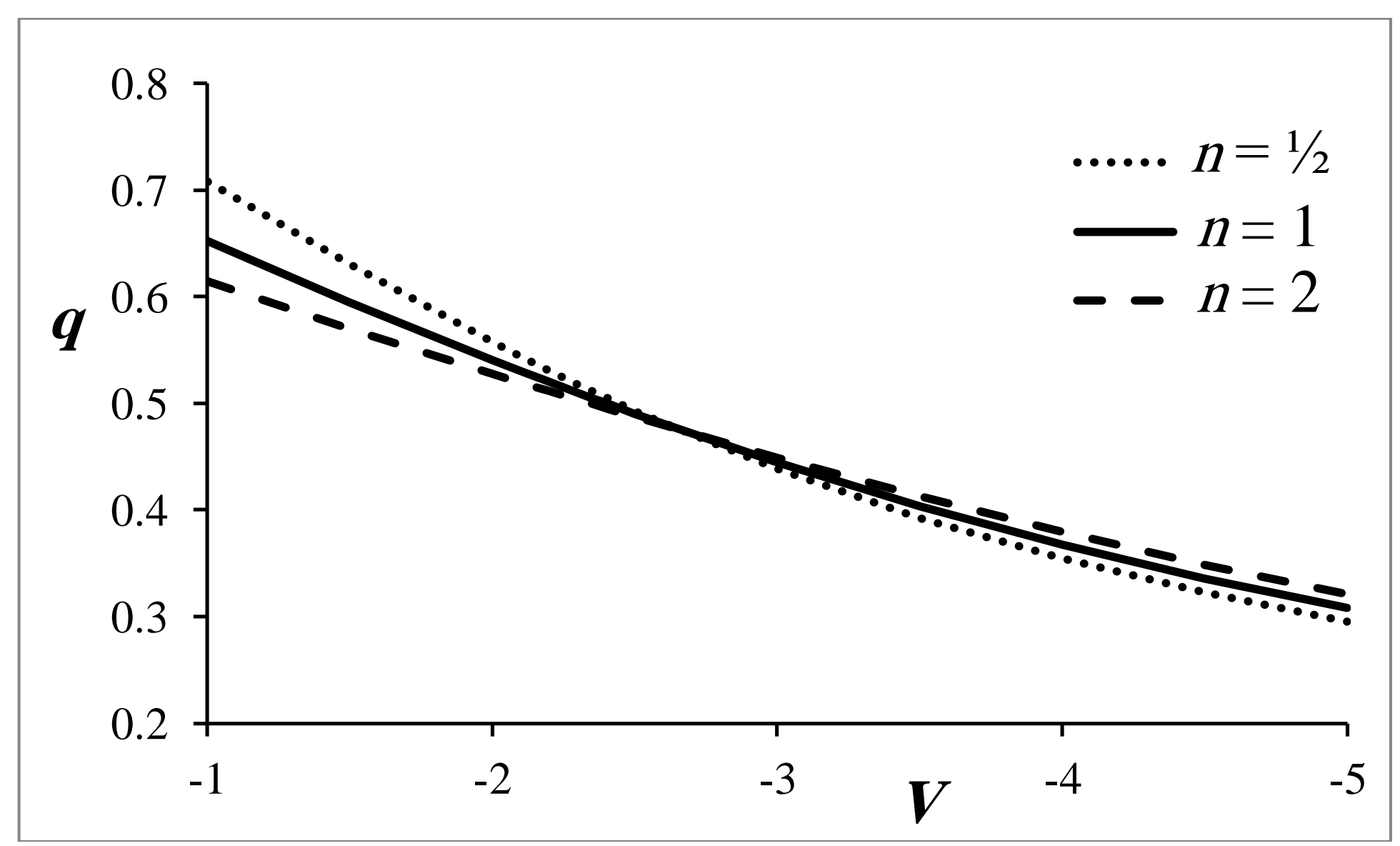

Fig. 4a: Flow rate vs. cross flow velocity, in Couette flow; $\pi$ =2.

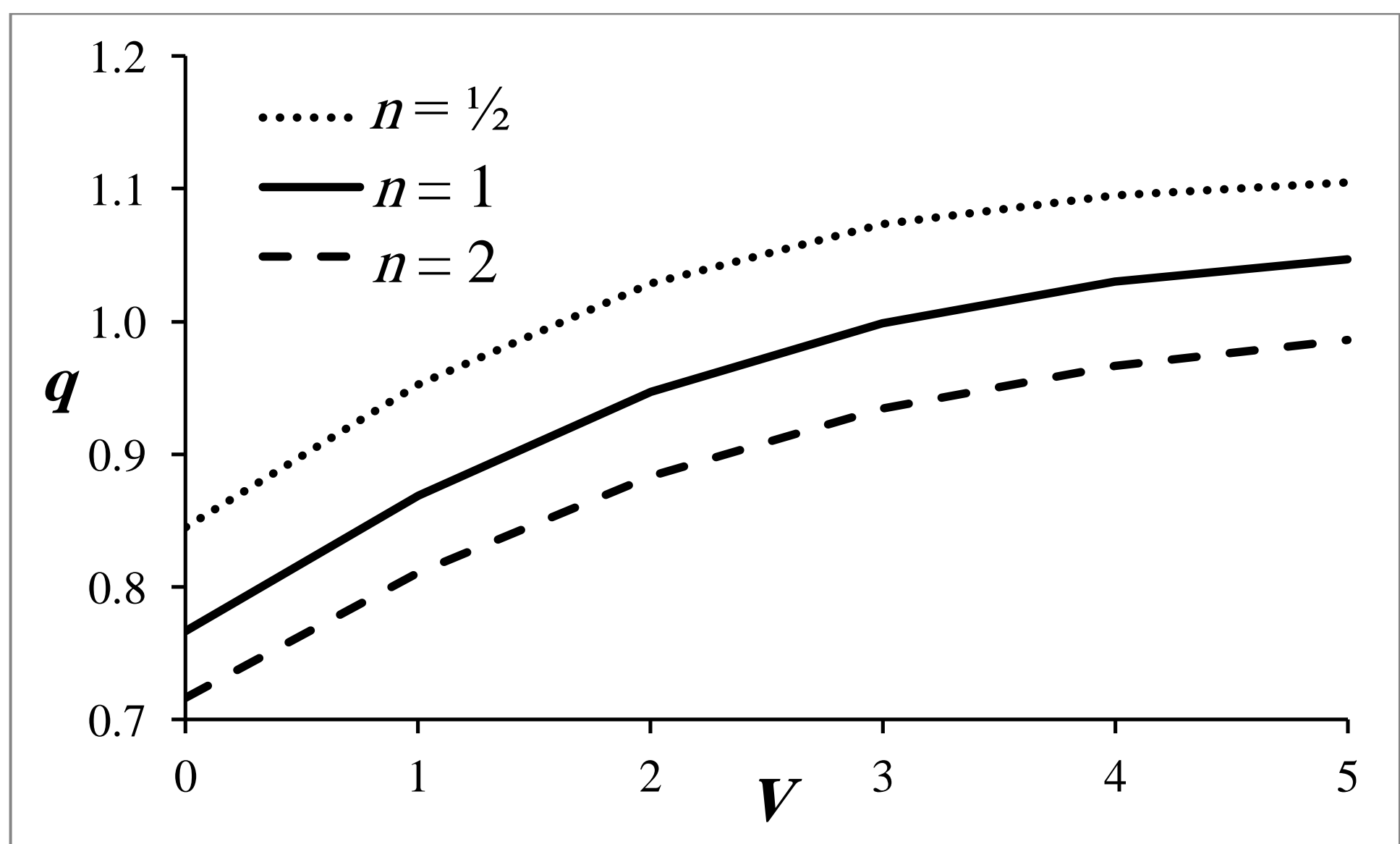

Fig. 4b: Flow rate vs. cross flow velocity, in Poiseuille flow; $\pi$ =2.

## 7. Conclusion

The effect of a uniform cross flow on the Couette-Poiseuille flow of power-law fluids with slip conditions has been studied. Depending on the strength of the cross flow, the flow can be of Couette type or of Poiseuille type.

Exact solutions have been obtained for illustrative cases of shear-thinning, Newtonian, and Shear-thickening fluids.

Sample results have been presented demonstrating velocity profiles and indicating variations of surface shears and flow rates, at representative values of the flow parameters.

Border-line values corresponding to linear velocity profiles or separating Couette-type flows from Poiseuille-type flows have been determined.

A case in which the shear-thickening power-law model fails has been identified.

## Appendix A: Limiting cases

Although the cases of $\pi$ =0 and/or $V$ =0 can be obtained as regular limits of the case $\pi$ >0 and $V$ ≠0, they are much easier when treated separately. The results are presented below without elaboration.

**The case of $\pi$ =0, $V$ ≠0**

The problem is unaltered under the transformation

$[V, \sigma_0, \sigma_1, u(y), \tau(y), q] \rightarrow [-V, \sigma_1, \sigma_0, w - u(1-y), \tau(1-y), w-q]$.

Therefore, only cases of $V$ >0 need to be considered.

The flow is strictly of Couette type (Appendix B, Note #6), and the velocity profile is convex (Appendix B, Note #7).

When $n$ =1, $\tau_0$ <0 is given by

$$\tau_0 = \frac{-Vw}{(e^V - 1) + V(\sigma_0 + \sigma_1 e^V)} \tag{A1}$$

while

$$\tau = \tau_0 e^{Vy} \tag{A2}$$

and

$$q = w + [\sigma_0 + (e^V - 1 - V)/V^2]\tau_0 \tag{A3}$$

When $n \neq 1$, $\tau_0$ satisfies

$$(1 + V\sigma_1)[(-\tau_0)^{\frac{n-1}{n}} + \frac{n-1}{n}V]^{\frac{n}{n-1}} + (1 - V\sigma_0)\tau_0 - Vw = 0 \tag{A4}$$

while

$$\tau = -[(-\tau_0)^{\frac{n-1}{n}} + \frac{n-1}{n}Vy]^{\frac{n}{n-1}} \tag{A5}$$

For $n$ =½, $\tau_0$ <0 is given by

$$\tau_0 = \frac{-2w}{(\sigma_0 + \sigma_1 + Vw) + \sqrt{(\sigma_0 + \sigma_1 + Vw)^2 + 4w(1 - V\sigma_0)}} \tag{A6}$$

which is always real (Appendix B, Note # 8) while

$$\tau = -[(-\tau_0)^{-1} - Vy]^{-1} \tag{A7}$$

and

$$q = w + \sigma_0\tau_0 - \frac{\tau_0}{V} + \frac{1}{V^2}\ln(1 + V\sigma_0) \tag{A8}$$

For $n$ =2, $\tau_0 \le 0$ is given by

$$\tau_0 = -\left\{-1 + \sqrt{1 + \left[\frac{4w}{(1 + V\sigma_1)} - V\right]\left[\frac{(\sigma_0 + \sigma_1)}{(1 + V\sigma_1)}\right]}\right\}^2 \Bigg/ \left[2\frac{(\sigma_0 + \sigma_1)}{(1 + V\sigma_1)}\right]^2 \tag{A9}$$

while

$$\tau = -[\sqrt{-\tau_0} + \tfrac{1}{2}Vy]^2 \tag{A10}$$

and

$$q = w + \sigma_0\tau_0 - \tfrac{1}{2}\sqrt{-\tau_0} - \tfrac{1}{12}V \tag{A11}$$

In this case of $n$ =2, $\tau_0$ reaches its top value of zero at a finite value of $V$ (Appendix B, Note #9),

$$V_F = \frac{-1 + \sqrt{1 + 16\sigma_1 w}}{2\sigma_1} \tag{A12}$$

indicating failure of the model, for $V > V_F$. In contrast, the cases of $n$ =½ and $n$ =1 allow $V$ to approach infinity (Appendix B, Note #10), with $\tau$ approaching zero everywhere except at $y = 1$, where $\tau_1 = -2w/(\sigma_0 + \sigma_1)$ and $\tau_1 = -w/\sigma_1$, respectively.

In all three cases $n$ =½, 1 and 2, $u$ is given by

$$u = w + \sigma_0\tau_0 + (\tau - \tau_0)/V \tag{A13}$$

**The case of $\pi$ =0, $V$ =0**

The flow is strictly of Couette type, and the velocity profile is linear; i.e. $\tau$ <0 is constant, satisfying (Appendix B, Note #12)

$$(\sigma_0 + \sigma_1)(-\tau) + (-\tau)^{1/n} - w = 0 \tag{A14}$$

and

$$u = w + \sigma_0\tau - (-\tau)^{1/n}y \tag{A15}$$

The flow rate is given by

$$q = [w + (\sigma_0 - \sigma_1)\tau]/2 \tag{A16}$$

For $n$ =½,

$$\tau = -\tfrac{1}{2}[-(\sigma_0 + \sigma_1) + \sqrt{(\sigma_0 + \sigma_1)^2 + 4w}] \tag{A17}$$

For $n$ =1,

$$\tau = \frac{-w}{\sigma_0 + \sigma_1 + 1} \tag{A18}$$

For $n$ =2,

$$\tau = -[\frac{-1 + \sqrt{1 + 4w(\sigma_0 + \sigma_1)}}{2(\sigma_0 + \sigma_1)}]^2 \tag{A19}$$

**The case of $\pi$ >0, $V$ =0**

The velocity profile, in this case, is convex. For small $\pi$, the flow is of Couette type. When $\pi$ exceeds a finite value $\pi_F$ given by

$$\frac{n}{n+1}\pi_F^{1/n} + \sigma_1\pi_F - w = 0 \tag{A20}$$

at which $\tau_0$ =0, the flow becomes of forward Poiseuille type (Appendix B, Note #13).

In cases of Couette flow, $\tau_0 \leq 0$ satisfies

$$\frac{n}{n+1}\frac{(\pi - \tau_0)^{\frac{n+1}{n}} - (-\tau_0)^{\frac{n+1}{n}}}{\pi} + \sigma_1(\pi - \tau_0) + \sigma_0(-\tau_0) - w = 0 \tag{A21}$$

with

$$\tau = \tau_0 - \pi y \tag{A22}$$

and

$$u = w + \sigma_0\tau_0 - \frac{n}{n+1}\frac{(\pi y - \tau_0)^{\frac{n+1}{n}} - (-\tau_0)^{\frac{n+1}{n}}}{\pi} \tag{A23}$$

The flow rate is given by

$$q = w + \sigma_0\tau_0 + \frac{n}{n+1}[\frac{(-\tau_0)^{\frac{n+1}{n}}}{\pi} - \frac{n}{2n+1}\frac{(\pi - \tau_0)^{\frac{2n+1}{n}} - (-\tau_0)^{\frac{2n+1}{n}}}{\pi^2}] \tag{A24}$$

For $n$ =½,

$$\tau_0 = [(\sigma_0 + \sigma_1 + \pi) - \sqrt{(\sigma_0 + \sigma_1 + \pi)^2 - 4(\pi^2/3 + \sigma_1\pi - w)}]/2 \tag{A25}$$

and

$$\pi_F = (\sqrt{9\sigma_1{}^2 + 12w} - 3\sigma_1)/2 \tag{A26}$$

For $n$ =1,

$$\tau_0 = \frac{\left(\sigma_1 + \frac{1}{2}\right)\pi - w}{\sigma_0 + \sigma_1 + 1} \tag{A27}$$

and

$$\pi_F = \frac{w}{\sigma_1 + \frac{1}{2}} \tag{A28}$$

For $n$ =2, $\tau_0$ satisfies

$$\frac{2}{3}\frac{(\pi - \tau_0)^{3/2} - (-\tau_0)^{3/2}}{\pi} + \sigma_1(\pi - \tau_0) + \sigma_0(-\tau_0) - w = 0 \tag{A29}$$

and

$$\pi_F = \left(\frac{-1 + \sqrt{1 + 9w\sigma_1}}{3\sigma_1}\right)^2 \tag{A30}$$

In cases of forward Poiseuille flow, $\tau_0$ >0 satisfies

$$w + \sigma_0\tau_0 - \frac{n}{n+1}\frac{(\pi - \tau_0)^{\frac{n+1}{n}} - \tau_0^{\frac{n+1}{n}}}{\pi} - \sigma_1(\pi - \tau_0) = 0 \tag{A31}$$

while

$$\tau = \tau_0 - \pi y \tag{A32}$$

with

$$y^* = \tau_0/\pi \tag{A33}$$

and

$$u^* = w + \sigma_0\tau_0 + \frac{n}{n+1}\frac{\tau_0^{\frac{n+1}{n}}}{\pi} \tag{A34}$$

When $0 \leq y \leq y^*$,

$$u = u^* - \frac{n}{n+1}\frac{\tau^{\frac{n+1}{n}}}{\pi} \tag{A35}$$

When $y^* \leq y \leq 1$,

$$u = u^* - \frac{n}{n+1}\frac{(-\tau)^{\frac{n+1}{n}}}{\pi} \tag{A36}$$

The flow rate is given by

$$q = u^* - \frac{n}{n+1}\frac{n}{2n+1}\frac{(\pi-\tau_0)^{\frac{2n+1}{n}} + \tau_0^{\frac{2n+1}{n}}}{\pi^2} \tag{A37}$$

These above expressions coincide with those of Hron et al. [6], when $w$ =0.

The cases of $n$ =½, 1 and 2 can be readily attained. In particular, for $n$ =1,

$$\tau_0 = \frac{(\sigma_1 + \frac{1}{2})\pi - w}{\sigma_0 + \sigma_1 + 1} \tag{A38}$$

and

$$u = w + \sigma_0\tau_0 + \tau_0 y - \tfrac{1}{2}\pi y^2 \tag{A39}$$

Substituting Eq. (A38) in Eq. (A39) results in an expression for $u$, which - properly rescaled- coincides with the expression of Matthews and Hill [5], when $w$ =0 and $\sigma_1 = \sigma_0$.

## Appendix B: Proofs

We give proofs of some statements and relations appearing in the main text and in Appendix A.

------

Note on signs:

$$\frac{du}{dy} > 0 \overset{(4)}{\Longrightarrow} \tau = \left(\frac{du}{dy}\right)^n \Longrightarrow \frac{du}{dy} = \tau^{\frac{1}{n}} \;\&\; \frac{d\tau}{dy} = n\left(\frac{du}{dy}\right)^{n-1}\frac{d^2u}{dy^2}$$

$$\frac{du}{dy} < 0 \overset{(4)}{\Longrightarrow} \tau = -\left(-\frac{du}{dy}\right)^n \Longrightarrow \frac{du}{dy} = -(-\tau)^{\frac{1}{n}} \;\&\; \frac{d\tau}{dy} = n\left(-\frac{du}{dy}\right)^{n-1}\frac{d^2u}{dy^2}$$

Conclusion:

$$\mathrm{sgn}\left(\frac{d\tau}{dy}\right) = \mathrm{sgn}\left(\frac{d^2u}{dy^2}\right) \tag{B1}$$

------

Note #1: For $\pi \geq 0$, the flow cannot be of backward Poiseuille type.

Proof:

For a backward Poiseuille-type flow, $\frac{d^2u}{dy^2} > 0$ for $0 < y < 1$. Moreover, there is $0 < y^* < 1$ such that $\frac{du}{dy} < 0$ for $0 \leq y < y^*$ and $\frac{du}{dy} > 0$ for $y^* < y \leq 1$. Let $V > 0$, then for $0 \leq y < y^*$, Eq. (1) gives $\frac{d\tau}{dy} = V\frac{du}{dy} - \pi < 0 \overset{(B1)}{\Longrightarrow} \frac{d^2u}{dy^2} < 0 \Longrightarrow$ contradiction. Let $V < 0$, then for $y^* < y \leq 1$, Eq. (1) gives $\frac{d\tau}{dy} = V\frac{du}{dy} - \pi < 0 \overset{(B1)}{\Longrightarrow} \frac{d^2u}{dy^2} < 0 \Longrightarrow$ contradiction. Let $V = 0$, then Eq. (1) gives $\frac{d\tau}{dy} = -\pi \leq 0 \overset{(B1)}{\Longrightarrow} \frac{d^2u}{dy^2} \leq 0 \Longrightarrow$ contradiction.

------

Note #2: Couette-type flows are characterized by $\frac{du}{dy} < 0$ for $0 < y < 1$.

Regardless of the value of $V$ or $\pi$ the following is true.

------

The velocity $u$ cannot be uniform.

Proof: Let $\frac{du}{dy} = 0$ for $0 \leq y \leq 1$. Then, $u_0 = u_1$. Eq. (4) gives $\tau = 0$ for $0 \leq y \leq 1$. Eq. (2) −Eq. (3) gives $u_0 - u_1 = w \neq 0 \Longrightarrow$ contradiction.

------

The velocity $u$ cannot be monotonically increasing.

Proof: Let $\frac{du}{dy} > 0$ for $0 < y < 1$ ($\frac{du}{dy}\Big|_{y=0} \geq 0$ and $\frac{du}{dy}\Big|_{y=1} \geq 0$). Then, $y' < y" \Longrightarrow u' < u"$. In particular, $u_0 < u_1$. Eq. (4) gives $\tau = (\frac{du}{dy})^n > 0$ for $0 < y < 1$ ($\tau_0 \geq 0$ and $\tau_1 \geq 0$). Eq. (2) −Eq. (3) gives $u_0 - u_1 = w + \sigma_0\tau_0 + \sigma_1\tau_1 > 0$; i.e. $u_0 > u_1 \Longrightarrow$ contradiction.

------

Conclusion:

The velocity $u$ is monotonically decreasing; i.e. $\frac{du}{dy} < 0$ for $0 < y < 1$ ($\frac{du}{dy}\Big|_{y=0} \leq 0$ and $\frac{du}{dy}\Big|_{y=1} \leq 0$). Therefore, $y' < y" \Longrightarrow u' > u"$. In particular, $u_0 > u_1$. Eq. (4) gives $\tau = -(-\frac{du}{dy})^n < 0$ for $0 < y < 1$ ($\tau_0 \leq 0$ and $\tau_1 \leq 0$). Then, Eq. (2) gives $u_0 \leq w$ and Eq. (3) gives $u_1 \geq 0$. Therefore, $w \geq u_0 > u_1 \geq 0$. Moreover, Eq. (2) gives $0 \geq \tau_0 > -\frac{w}{\sigma_0} \Longrightarrow 0 \geq \frac{du}{dy}\Big|_{y=0} > -(\frac{w}{\sigma_0})^{\frac{1}{n}}$.

------

Consider Couette-type flows. For $\pi > 0$ the following is true.

------

Note #3: When $V\frac{du}{dy} < \pi$ the velocity profile is convex, with its slope $\frac{du}{dy}$ monotonically decreasing ($\left|\frac{du}{dy}\right|$ monotonically increasing) with increasing $y$. (This is obviously true when $V \geq 0$, since $V\frac{du}{dy} < 0 < \pi$.)

Proof:

$$V\frac{du}{dy} < \pi \overset{(1)}{\Longleftrightarrow} \frac{d\tau}{dy} = V\frac{du}{dy} - \pi < 0 \overset{(B1)}{\Longleftrightarrow} \frac{d^2u}{dy^2} < 0 \Longrightarrow \text{convex velocity profile.}$$

------

Note #4: When $V\frac{du}{dy} = \pi$ the velocity profile is linear. In this case, the value of $V = V_L < 0$, is given by

$$(\sigma_0 + \sigma_1)(-\frac{\pi}{V_L})^n + \left(-\frac{\pi}{V_L}\right) - w = 0$$

Proof:

$V\frac{du}{dy} = \pi \overset{(1)}{\Longleftrightarrow} \frac{d\tau}{dy} = V\frac{du}{dy} - \pi = 0 \Longrightarrow \tau = \tau_L$ a constant $\overset{(4)}{\Longleftrightarrow}$ constant $\frac{du}{dy} = \frac{du}{dy}\Big|_L \Longrightarrow$ linear velocity profile. The corresponding value of $V = V_L < 0$ is such that $\frac{du}{dy}\Big|_L = \frac{\pi}{V_L}$, which integrates to give $u_{1L} - u_{0L} = \frac{\pi}{V_L}$. On the other hand, Eq. (2) −Eq. (3) gives $u_{0L} - u_{1L} - (\sigma_0 + \sigma_1)\tau_L = w$. Thus, noting that $\tau_L \overset{(4)}{\Longrightarrow} -(-\frac{\pi}{V_L})^n$, we get $(\sigma_0 + \sigma_1)\left(-\frac{\pi}{V_L}\right)^n + \left(-\frac{\pi}{V_L}\right) - w = 0$, or, alternatively, $(\sigma_0 + \sigma_1)(-\tau_L) + (-\tau_L)^{1/n} - w = 0$.

------

Note #5: When $V < V_L < 0$ and $\frac{du}{dy} < 0$ are such that $V\frac{du}{dy} > \pi > 0$, the velocity profile is concave, with its slope $\frac{du}{dy}$ monotonically increasing ($\left|\frac{du}{dy}\right|$ monotonically decreasing) with increasing $y$.

Proof:

$$V\frac{du}{dy} > \pi \overset{(1)}{\Longrightarrow} \frac{d\tau}{dy} = V\frac{du}{dy} - \pi > 0 \overset{(\mathrm{B1})}{\Longrightarrow} \frac{d^2u}{dy^2} > 0 \Longrightarrow \text{concave velocity profile.}$$

------

When $\pi = 0$ and $V > 0$, the following is true.

------

Note #6: The flow is strictly of Couette type.

Proof:

Let the flow be of forward Poiseuille type; i.e. $\frac{d^2u}{dy^2} < 0$ for $0 < y < 1$. Moreover, there is $0 < y^* < 1$ such that $\frac{du}{dy} > 0$ for $0 \leq y < y^*$. Then, for $0 \leq y < y^*$, with $V > 0$, Eq. (1) gives $\frac{d\tau}{dy} = V\frac{du}{dy} > 0 \overset{(\mathrm{B1})}{\Longrightarrow} \frac{d^2u}{dy^2} > 0 \Longrightarrow$ contradiction.

------

Note #7: The velocity profile is convex.

Proof:

For Couette-type flows, $\frac{du}{dy} < 0$ for $0 < y < 1$. Then, Eq. (1) gives $\frac{d\tau}{dy} = V\frac{du}{dy} < 0 \overset{(\mathrm{B1})}{\Longrightarrow} \frac{d^2u}{dy^2} < 0 \Longrightarrow$ convex velocity profile,

------

Note #8: For $n$ =½, $\tau_0 < 0$ given by Eq. (A6) is always real.

Proof:

The discriminant of the square root in Eq. (A6) is

$$(\sigma_0 + \sigma_1 + Vw)^2 + 4w(1 - V\sigma_0) = (\sigma_0)^2 + (\sigma_1)^2 + (Vw)^2 + 2\sigma_0\sigma_1 + 2\sigma_0 Vw + 2\sigma_1 Vw + 4w - 4wV\sigma_1 = (\sigma_0)^2 + (\sigma_1 - Vw)^2 + 2\sigma_0\sigma_1 + 2\sigma_0 Vw + 4w > 0.$$

------

Note #9: For $n$ =2, $\tau_0 \leq 0$ given by Eq. (A9) reaches its top value of zero at a finite value of $V = V_F$.

Proof:

See Note #11 below.

------

Note #10: The cases of $n$ =½ and $n$ =1 allow $V$ to approach infinity, with $\tau$ approaching zero everywhere except at $y = 1$, where $\tau_1 = -w/\sigma_1$, so that $u = w$ for $0 \leq y \leq 1$.

Proof:

See Note #11 below, and take limits as $V \sim \infty$ of $\tau_0$, $\tau_1$ and $\tau$.

------

Note #11: Generalization: In the case of $n > 1$, $\tau_0$ reaches its top value of zero at a finite value of $V = V_F$. In contrast, the cases of $n < 1$ and $n$ =1 allow $V$ to approach infinity.

Proof:

$$\pi = 0 \overset{(8,9_)}{\overbrace{\Longrightarrow}} \lim_{\alpha \to 0} \int_{\alpha}^{\frac{V_F w}{1+V_F \sigma_1}} \frac{d\hat{\tau}}{\hat{\tau}^{1/n}} = V_F$$

which gives, upon performing the integration, rearranging, and invoking the limit,

$$n < 1: [(\frac{V_F w}{1 + V_F \sigma_1})^{-\frac{1-n}{n}} + (\frac{1-n}{n})V_F]^{-1} = 0$$

$$n = 1: (\frac{V_F w}{1 + V_F \sigma_1})e^{-V_F} = 0$$

$$n > 1: (\frac{V_F w}{1 + V_F \sigma_1})^{\frac{n-1}{n}} - (\frac{n-1}{n})V_F = 0$$

In the first two cases, the equation can be satisfied only when $V_F \sim \infty$.

In the third case, the equation is satisfied at a finite value of $V_F$ given by

$$(\frac{V_F w}{1 + V_F \sigma_1})^{\frac{n-1}{n}} = (\frac{n-1}{n})V_F$$

In particular, when $n = 2$, $V_F$ satisfies

$$(\frac{V_F w}{1 + V_F \sigma_1})^{\frac{1}{2}} = \frac{1}{2}V_F$$

which can be manipulated to the quadratic equation [inferred also from Eq. (A9)]

$$\sigma_1 {V_F}^2 + V_F - 4w = 0$$

with the positive solution

$$V_F = \frac{-1 + \sqrt{1 + 16\sigma_1 w}}{2\sigma_1}$$

which is Eq. (A12).

------

Note #12: In the case of $\pi$ =0 and $V$ =0, the flow is strictly of Couette type, and the velocity profile is linear; i.e. $\tau$ <0 is constant, satisfying

$$(\sigma_0 + \sigma_1)(-\tau) + (-\tau)^{1/n} - w = 0$$

Proof:

Eq. (1) gives $\frac{d\tau}{dy} = 0 \Longrightarrow \tau$ is constant $\overset{(4)}{\Longleftrightarrow} \frac{du}{dy}$ is constant $\Longrightarrow$ Couette flow ($\frac{du}{dy} < 0$) with linear velocity profile. Integration of $\frac{du}{dy} = -(-\tau)^{\frac{1}{n}}$ gives $u_1 - u_0 = -(-\tau)^{\frac{1}{n}}$. Use of Eqs .(2) and (3) leads to $-\sigma_1\tau - (\sigma_0\tau + w) = -(-\tau)^{\frac{1}{n}} \Longrightarrow (\sigma_1 + \sigma_0)(-\tau) + (-\tau)^{\frac{1}{n}} - w = 0$.

------

Note #13: In the case of $\pi$ >0 and $V$ =0, the velocity profile is convex. For small $\pi$, the flow is of Couette type. When $\pi$ exceeds a finite value $\pi_F$ given by

$$\frac{n}{n+1}\pi_F^{1/n} + \sigma_1\pi_F - w = 0$$

at which $\tau_0$ =0, the flow becomes of forward Poiseuille type.

Proof:

Eq. (1) gives $\frac{d\tau}{dy} = -\pi < 0 \overset{(B1)}{\Longleftrightarrow} \frac{d^2u}{dy^2} < 0 \Longrightarrow$ convex velocity profile. The flow may be of Couette type or forward Poiseuille type, depending on the value of $\pi$.

For a Couette-type flow ( $\frac{du}{dy} < 0$ for $0 < y < 1$), integration of $\frac{d\tau}{dy} = -\pi$ gives $\tau - \tau_0 = -\pi y \Longrightarrow \frac{du}{dy} = -(\pi y - \tau_0)^{\frac{1}{n}}$ which integrates to $u_1 - u_0 = \frac{-(\pi-\tau_0)^{\frac{1}{n}+1}+(-\tau_0)^{\frac{1}{n}+1}}{\left(\frac{1}{n}+1\right)\pi}$. Use of Eqs. (2) and (3) with $\tau_1 = \tau_0 - \pi$ leads to the following equation for $\tau_0$

$$-\sigma_1(\tau_0 - \pi) - \sigma_0\tau_0 - w = \frac{-(\pi-\tau_0)^{\frac{1}{n}+1} + (-\tau_0)^{\frac{1}{n}+1}}{\left(\frac{1}{n}+1\right)\pi}$$

which differentiates with respect to $\pi$ to

$$\{[\sigma_1 + \sigma_0]\pi + (\pi - \tau_0)^{\frac{1}{n}} - (-\tau_0)^{\frac{1}{n}}\}\frac{d\tau_0}{d\pi} = [\sigma_1 + \sigma_0](-\tau_0) + 2\sigma_1\pi + (\pi - \tau_0)^{\frac{1}{n}} - w.$$

The value of $\pi = \pi_F$ at which $\tau_0 = 0$ is given by

$$\frac{n}{(n+1)}\pi^{\frac{1}{n}} + \sigma_1\pi - w = 0$$

at this value

$$\left.\frac{d\tau_0}{d\pi}\right|_{\tau_0=0} = \frac{2\sigma_1\pi + \pi^{\frac{1}{n}} - w}{[\sigma_1 + \sigma_0]\pi + \pi^{\frac{1}{n}}} = \frac{\sigma_1\pi + \pi^{\frac{1}{n}} - \frac{n}{(n+1)}\pi^{\frac{1}{n}}}{[\sigma_1 + \sigma_0]\pi + \pi^{\frac{1}{n}}} = \frac{\sigma_1\pi + \frac{1}{(n+1)}\pi^{\frac{1}{n}}}{[\sigma_1 + \sigma_0]\pi + \pi^{\frac{1}{n}}} > 0$$

Increasing $\pi$ above $\pi_F$ leads to $\tau_0 > 0$, in violation to the characteristic of Couette-type flows. Thus the flow becomes of Poiseuille type.

**Appendix C: Extension in accordance with Ref. [13]**

For fixed favorable pressure gradient $dp'/dx' < 0$ along the $x'$- axis (so that $\pi > 0$), Eqs. (1)-(5) are unaltered under the transformation

$$\{w, V, \sigma_0, \sigma_1, u(y), \tau(y), q\} \rightarrow \{-w, -V, \sigma_1, \sigma_0, u(1-y) - w, -\tau(1-y), q - w\}$$

Therefore, we only need to consider cases of $V \geq 0$, allowing $w$ to vary. This is the choice of Ref. [13].

When $V > 0$, the velocity profile can be one of the following.

(a) Convex of Couette type, with $\tau_1 < \tau_0 \leq 0$.
(b) Convex of forward Poiseuille type, with $(\pi/V)^n > \tau_0 > 0 > \tau_1$.
(c) Convex of Couette type, with $(\pi/V)^n > \tau_0 > \tau_1 \geq 0$.
(d) Linear of Couette type, with $\tau_1 = \tau_0 = \tau_L = (\pi/V)^n$.
(e) Concave of Couette type, with $\tau_1 > \tau_0 > (\pi/V)^n > 0$.

As indicated in the main text the flow cannot be of backward Poiseuille type.

In order to write the indicated inequalities, we have utilized: $d\tau/dy < 0$ for the convex profiles (a)-(c), $d\tau/dy > 0$ for the concave profile (e), and obviously $d\tau/dy = 0$ for the linear profile (d).

Lemma: In all of cases (a) to (e),

$$\frac{d\tau_0}{dw} < 0 \tag{C1}$$

Proof:

Equation (8) differentiates with respect to $w$ to give

$$\frac{d\tau_1}{dw} - L\frac{d\tau_0}{dw} = R, L = \frac{1 - V\sigma_0}{1 + V\sigma_1} < 1, R = -\frac{V}{1 + V\sigma_1} < 0 \tag{C2}$$

(a) Equation (10) applies and differentiates with respect to $w$ to give

$$\frac{d\tau_1}{dw} = L_a\frac{d\tau_0}{dw}, L_a = \frac{(-\tau_1)^{1/n} + \pi/V}{(-\tau_0)^{1/n} + \pi/V} > 1 \tag{C3a}$$

Substituting in (C2) we get

$$[L_a - L]\frac{d\tau_0}{dw} = R \tag{C4a}$$

(b) Equation (30) applies and differentiates with respect to $w$ to give

$$\frac{d\tau_1}{dw} = L_b \frac{d\tau_0}{dw}, L_b = \frac{(-\tau_1)^{1/n} + \pi/V}{\pi/V - \tau_0{}^{1/n}} > 1 \tag{C3b}$$

Substituting in (C2) we get

$$[L_b - L] \frac{d\tau_0}{dw} = R \tag{C4b}$$

(c) Equation ($9_+$), evaluated at $y = 1$ with $i = 0$, applies and differentiates with respect to $w$ to give

$$\frac{d\tau_1}{dw} = L_c \frac{d\tau_0}{dw}, L_c = \frac{\tau_1{}^{1/n} - \pi/V}{\tau_0{}^{1/n} - \pi/V} > 1 \tag{C3c}$$

Substituting in (C2) we get

$$[L_c - L] \frac{d\tau_0}{dw} = R \tag{C4c}$$

(d) Since $\tau_1 = \tau_0 = \tau_L$, we can write

$$\frac{d\tau_1}{dw} = L_d \frac{d\tau_0}{dw}, L_d = 1 \tag{C3d}$$

Substituting in (C2) we get

$$[L_d - L] \frac{d\tau_0}{dw} = R \tag{C4d}$$

(e) Equation ($9_+$), evaluated at $y = 1$ with $i = 0$, applies and differentiates with respect to $w$ to give

$$\frac{d\tau_1}{dw} = L_e \frac{d\tau_0}{dw}, L_e = \frac{\tau_1{}^{1/n} - \pi/V}{\tau_0{}^{1/n} - \pi/V} > 1 \tag{C3e}$$

Substituting in (C2) we get

$$[L_e - L] \frac{d\tau_0}{dw} = R \tag{C4e}$$

Equations (C3) and (C4) combine, respectively, as

$$\frac{d\tau_1}{dw} = L_* \frac{d\tau_0}{dw}, L_* \geq 1 \tag{C5}$$

$$[L_* - L] \frac{d\tau_0}{dw} = R \tag{C6}$$

In (C6), $L_* \geq 1$ and $L < 1$, therefore the bracket in the left hand side is positive, while the right hand side $R$ is negative. Hence (1).

Corollaries:

$$\frac{d\tau_1}{dw} < 0 \tag{C7}$$

$$\frac{du_1}{dw} > 0 \tag{C8}$$

$$\frac{du_0}{dw} > 0 \tag{C9}$$

$$\frac{du_1}{dw} \leq \frac{du_0}{dw} \tag{C10}$$

Proof

(C7) follows from (C1) and (C5).

Differentiate (3) with respect to $w$ to get

$$\frac{du_1}{dw} = -\sigma_1 \frac{d\tau_1}{dw} \tag{C11}$$

Then (C8) follows from (C7).

Differentiate (2) with respect to $w$ to get

$$\frac{du_0}{dw} = 1 + \sigma_0 \frac{d\tau_0}{dw} \tag{C12}$$

Multiplication by $[L_* - L] > 0$ and use of (C6) result in

$$[L_* - L]\frac{du_0}{dw} = [L_* - L] + \sigma_0 R = L_* - \frac{1}{1 + V\sigma_1} > 0$$

since $L_* \geq 1$ and $\frac{1}{1+V\sigma_1} < 1$. Hence (C9).

Subtraction of (C11) from (C12) and use of (C5) give

$$\frac{du_0}{dw} - \frac{du_1}{dw} = 1 + \frac{d\tau_0}{dw}(\sigma_0 + \sigma_1 L_*)$$

Multiplication by $[L_* - L] > 0$ and use of (C6) result in

$$[L_* - L](\frac{du_0}{dw} - \frac{du_1}{dw}) = [L_* - L] + R(\sigma_0 + \sigma_1 L_*) = [L_* - 1]\frac{1}{1 + V\sigma_1} \geq 0$$

Hence (C10). The equality holds in case (d) only.

As $w$ decreases, $\tau_0$ increases and so does $\tau_1$ but at a faster rate as (C1), (C7) and (C5) indicate, while $u_0$ decreases and so does $u_1$ but at a faster rate as (C9), (C8) and (C10) indicate, respectively.

Within the Couette flow of convex profile {case (a)}, as $w$ increases to $\infty$, $\tau_1 < 0$ decreases approaching $-\infty$ with $\tau_0 < 0$ approaching $-\infty$ or a lowest possible value $\tau_{0low}$. On the other hand, as $w$ decreases $\tau_0$ increases reaching the value of 0, when $w$ reaches the value $w = w_{F0}$. As $w$ decreases below $w_{F0}$, the flow becomes of forward Poiseuille type {case (b)}. Within this case, as $w$ decreases $\tau_1 < 0$ increases approaching the value of 0, when $w$ approaches the value $w = w_{F1}$, and the flow becomes of Couette type with convex profile {case (c)}. As $w$ decreases further reaching a value $w = w_L$, the profile becomes linear {case(d)}. Below this value of $w$ the flow is of Couette type with concave profile {case (e)}. As $w$ decreases to $-\infty$, $\tau_1 > 0$ increases approaching $\infty$ with $\tau_0 > 0$ approaching $\infty$ or a highest possible value $\tau_{0high}$.

For large $w > 0$, $\tau_0 < 0$ and $u_0 = \sigma_0\tau_0 + w > 0$. As $w$ decreases $u_0$ decreases reaching the value of zero when $w = w_0$, then turning negative. $u_1 = -\sigma_1\tau_1$ decreases, as well, reaching the value of zero when $w = w_{F1} < w_0$.

Following the same steps given in Note #4 of Appendix B, we can prove that

$$w_L = -[(\sigma_0 + \sigma_1)(\frac{\pi}{V})^n + \frac{\pi}{V}] \tag{C5}$$

The other border-line values $w_{F0}$ and $w_{F1}$ (or the equations defining them), as well as the behavior of $\tau_1$ and $\tau_0$ as $w \sim \pm\infty$ (or the values of $\tau_{0low}$ and $\tau_{0high}$) are given below in the cases of $n$ = ½, 1 and 2.

**The case of shear-thinning fluid ($n$ = ½)**

$$w \sim \infty: \tau_1 \sim -\frac{Vw}{1+V\sigma_1}, \tau_{0low} \sim -\frac{\sqrt{\pi/V}}{\tan\sqrt{\pi V}}$$

$$w_{F0} = \frac{\pi}{V}[(1+V\sigma_1)\frac{\tan\sqrt{\pi V}}{\sqrt{\pi V}} - 1]: \tau_1 = -\sqrt{\pi/V}\tan\sqrt{\pi V}, \tau_0 = 0$$

$$w_L = -[(\sigma_0 + \sigma_1)\sqrt{\pi/V} + \pi/V]: \tau_L = \sqrt{\pi/V}$$

$$w_{F1} = -\frac{\pi}{V}[1 - (1 - V\sigma_0)\frac{\tanh\sqrt{\pi V}}{\sqrt{\pi V}}]: \tau_1 = 0, \tau_0 = \sqrt{\pi/V}\tanh\sqrt{\pi V}$$

$$w = -\infty: \tau_1 \sim -\frac{Vw}{1+V\sigma_1}, \tau_{0high} \sim \frac{\sqrt{\pi/V}}{\tanh\sqrt{\pi V}}$$

**The case of Newtonian fluid ($n = 1$)**

$$w \sim \infty: \tau_0 \sim -\frac{Vw}{(1+V\sigma_1)e^V - (1-V\sigma_0)}, \tau_1 \sim -\frac{Vw}{(1+V\sigma_1) - (1-V\sigma_0)e^{-V}}$$

$$w_{F0} = \frac{\pi}{V}[(1+V\sigma_1)\frac{e^V - 1}{V} - 1]: \tau_0 = 0, \tau_1 = -\frac{\pi}{V}(e^V - 1)$$

$$w_L = -\frac{\pi}{V}(\sigma_0 + \sigma_1 + 1): \tau_L = \frac{\pi}{V}$$

$$w_{F1} = -\frac{\pi}{V}[1 - (1 - V\sigma_0)\frac{1-e^{-V}}{V}]: \tau_0 = \frac{\pi}{V}(1 - e^{-V}), \tau_1 = 0$$

$$w \sim -\infty: \tau_0 \sim -\frac{Vw}{(1+V\sigma_1)e^V - (1-V\sigma_0)}, \tau_1 \sim -\frac{Vw}{(1+V\sigma_1) - (1-V\sigma_0)e^{-V}}$$

**The case of shear-thickening fluid ($n = 2$)**

$w \sim \infty$: $\tau_0 \sim \frac{-w}{\sigma_1+\sigma_0}$, $\tau_1 \sim \frac{-w}{\sigma_1+\sigma_0}$

$\frac{V}{\pi}\sqrt{\frac{Vw_{F0}+\pi}{1+V\sigma_1}} - \ln(1+\frac{V}{\pi}\sqrt{\frac{Vw_{F0}+\pi}{1+V\sigma_1}}) = \frac{V^2}{2\pi}$: $\tau_0 = 0$, $\tau_1 = -\frac{Vw_{F0}+\pi}{1+V\sigma_1} < 0$

$w_L = -[(\sigma_0+\sigma_1)(\frac{\pi}{V})^2 + \frac{\pi}{V}]$: $\tau_L = (\frac{\pi}{V})^2$

$\ln(1-\frac{V}{\pi}\sqrt{\frac{Vw_{F1}+\pi}{1-V\sigma_0}})^{-1} - \frac{V}{\pi}\sqrt{\frac{Vw_{F1}+\pi}{1-V\sigma_0}} = \frac{V^2}{2\pi}$: $\tau_0 = \frac{Vw_{F1}+\pi}{1-V\sigma_0} > 0$, $\tau_1 = 0$

$w = -\infty$: $\tau_0 \sim \frac{-w}{\sigma_1+\sigma_0}$, $\tau_1 \sim \frac{-w}{\sigma_1+\sigma_0}$

Sample numerical results, when $\pi = 1$, $V = 1$, $\sigma_0 = \sigma_1 = 0.1$, are shown in Figs. C1 and Tables C1.

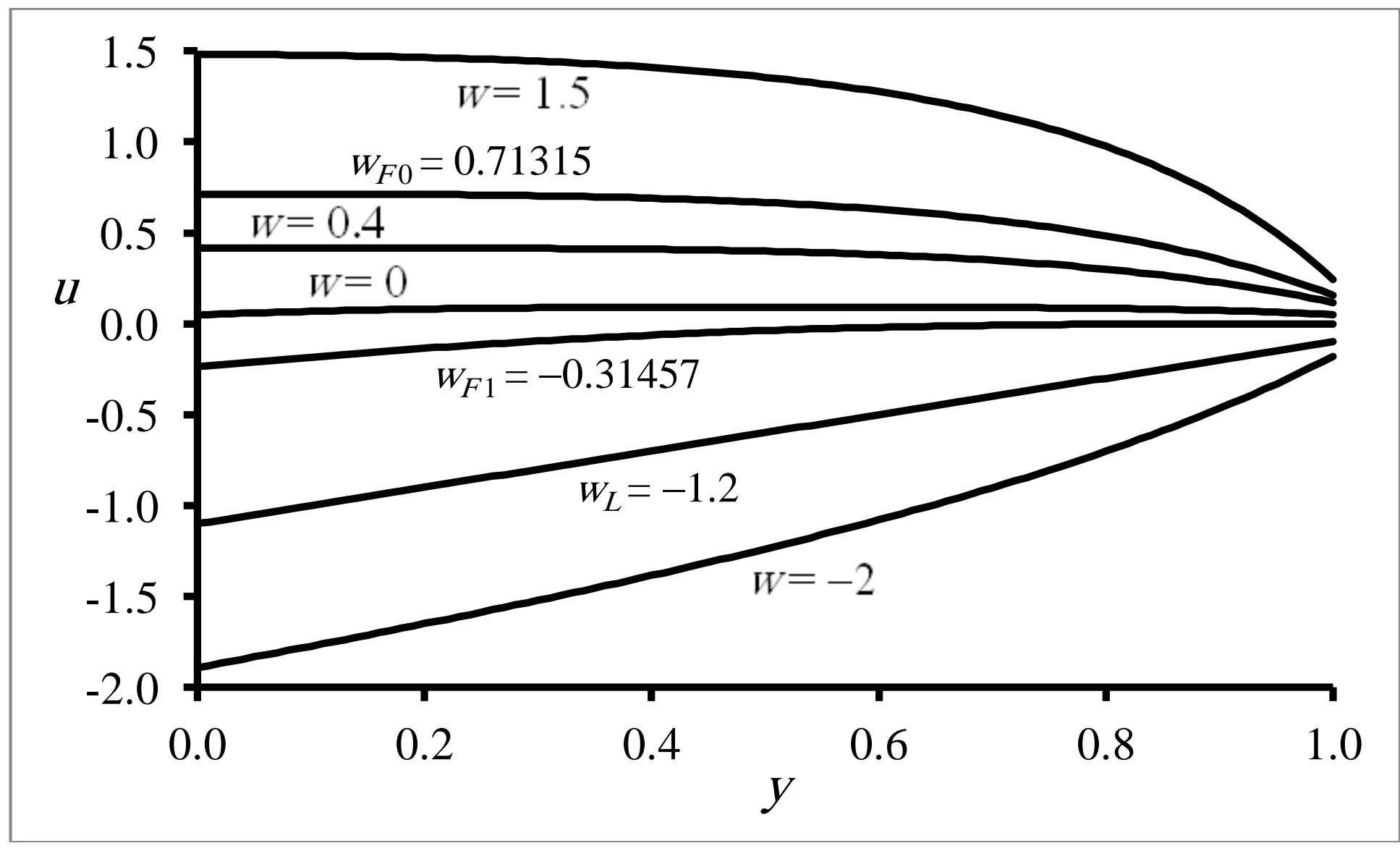

Fig. C1a: Velocity profiles for a shear-thinning fluid, $n = ½$; $\pi$ =1, $V$=1.

Table C1a: Sample results for $n = ½$, $\pi = 1$, $V = 1$, $\sigma_0 = \sigma_1 = 0.1$.

| Case | $w$ | $\tau_0$ | $\tau_1$ |
|---|---|---|---|
| $w > w_{F0}$ | 1.5 | -0.1810009773 | -2.4208189814 |
| $w = w_{F0}$ | 0.7131484971 | 0.0 | -1.5574077247 |
| $w < w_{F0}$ | 0.4 | 0.1407651401 | -1.1575557944 |
| $w = 0$ | 0.0 | 0.4805553543 | -0.5159092555 |
| $w = w_0$ | -0.0539628194 | 0.5396281941 | -0.4185198236 |
| $w = w_{F1}$ | -0.3145652596 | 0.7615941560 | 0.0 |
| $w = w_L$ | -1.2 | 1.0 | 1.0 |
| $w < w_L$ | -2.0 | 1.0798938609 | 1.7926404317 |

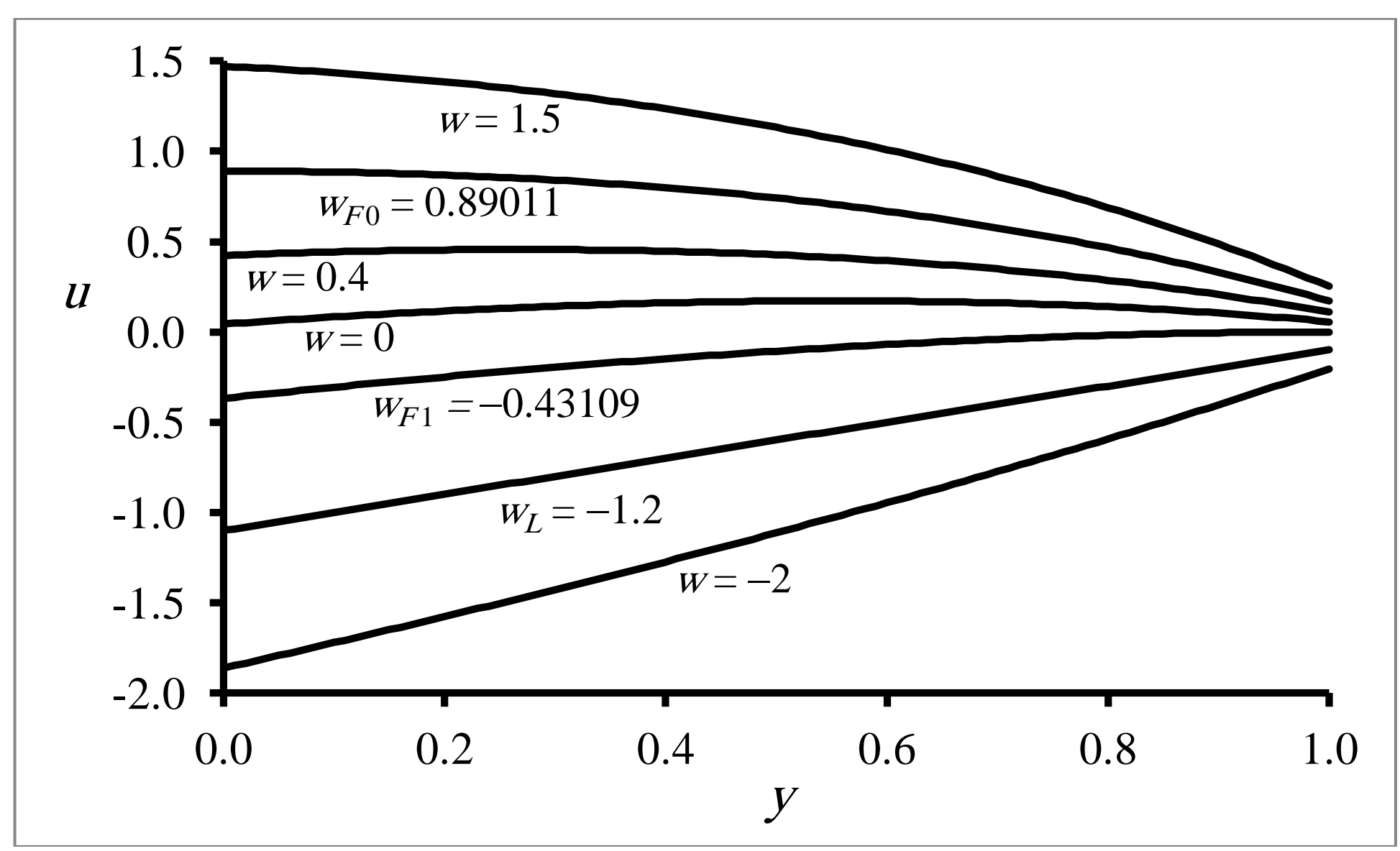

Fig. C1b: Velocity profiles for a shear-thinning fluid, $n = 1$; $\pi = 1$, $V$=1.

Table C1b: Sample results for $n = 1, \pi = 1, V = 1, \sigma_0 = \sigma_1 = 0.1$.

| Case | $w$ | $\tau_0$ | $\tau_1$ |
|---|---|---|---|
| $w > w_{F0}$ | 1.5 | -0.2917980324 | -2.5114711174 |
| $w = w_{F0}$ | 0.8901100113 | 0.0 | -1.7182818285 |
| $w < w_{F0}$ | 0.4 | 0.2344900549 | -1.0808717733 |
| $w = 0$ | 0.0 | 0.4258675412 | -0.5606538299 |
| $u_0 = 0$ | -0.0447266737 | 0.4472667371 | -0.5024847844 |
| $w = w_{F1}$ | -0.4310914971 | 0.6321205588 | 0.0 |
| $w = w_L$ | -1.2 | 1.0 | 1.0 |
| $w < w_L$ | -2.0 | 1.3827549725 | 2.0404358866 |

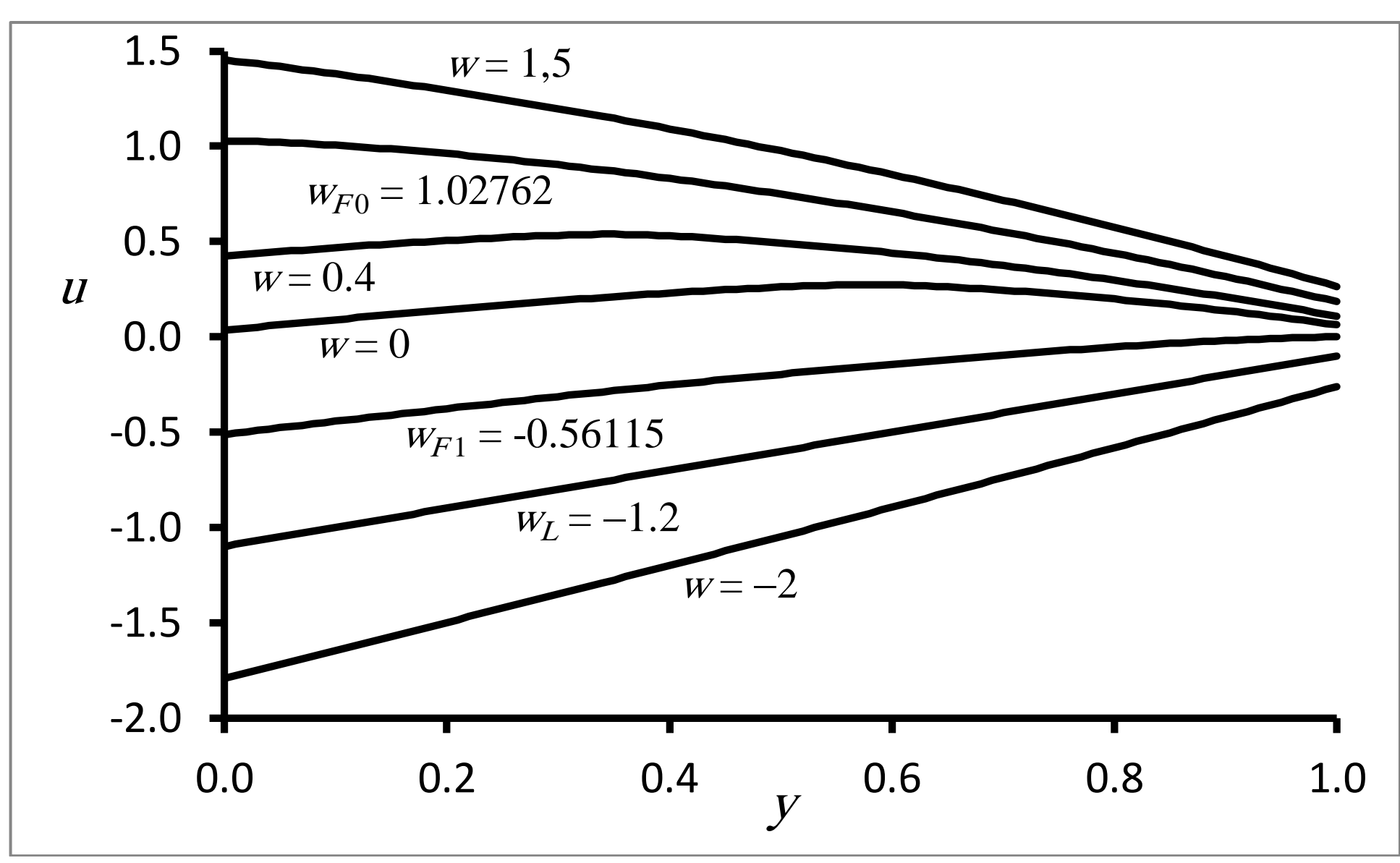

Fig. C1c: Velocity profiles for a shear-thinning fluid, $n$ =2; $\pi$ =1, $V$=1.

Table C1c: Sample results for $n = 2, \pi = 1, V = 1, \sigma_0 = \sigma_1 = 0.1$.

| Case | $w$ | $\tau_0$ | $\tau_1$ |
|---|---|---|---|
| $w > w_{F0}$ | 1.5 | -0.4673284067 | -2.6550868782 |
| $w = w_{F0}$ | 1.0276145461 | 0.0 | -1.8432859510 |
| $w < w_{F0}$ | 0.4 | 0.2282683515 | -1.0859622579 |
| $w = 0$ | 0.0 | 0.3384669691 | -0.6321633889 |
| $w = w_0$ | -0.0347534408 | 0.347534408 | -0.5931505378 |
| $w = w_{F1}$ | -0.5611514186 | 0.4876095348 | 0.0 |
| $w = w_L$ | -1.2 | 1.0 | 1.0 |
| $w < w_L$ | -2.0 | 2.0903055184 | 2.6193408787 |

When $V = 0$, the velocity profile can be one of the following.

(f) Convex of Couette type, with $\tau_1 < \tau_0 \leq 0$.
(g) Convex of forward Poiseuille type, with $\tau_0 > 0 > \tau_1$.
(h) Convex of Couette type, with $\tau_0 > \tau_1 \geq 0$.

and it can be proven that, in all three case,

$$\frac{d\tau_1}{dw} = \frac{d\tau_0}{dw} < 0$$

Thus, as $w$ decreases, we move from case (f) to case (g) and then to case (h). The following can be derived.

$w \sim \infty$: $\tau_0 \sim \frac{-w}{\sigma_0+\sigma_1}$, $\tau_1 \sim \frac{-w}{\sigma_0+\sigma_1}$

$w_{F0} = \sigma_1 \pi + \frac{n}{n+1} \pi^{\frac{1}{n}}$: $\tau_0 = 0, \tau_1 = -\pi$

$w_{F1} = -(\sigma_0 \pi + \frac{n}{n+1} \pi^{\frac{1}{n}})$: $\tau_0 = \pi, \tau_1 = 0$

$w \sim -\infty$: $\tau_0 \sim \frac{-w}{\sigma_0+\sigma_1}$, $\tau_1 \sim \frac{-w}{\sigma_0+\sigma_1}$

Sample numerical results, when $\pi = 1, V = 0, \sigma_0 = \sigma_1 = 0.1$, are shown in Tables C2 and Figs. C2.

Table C2a: Sample results for $n = ½, \pi = 1, V = 0, \sigma_0 = \sigma_1 = 0.1$.

| Case | $w$ | $\tau_0$ | $\tau_1$ | $u_0$ |
|---|---|---|---|---|
| $w > w_{F0}$ | 1.0 | -0.3626352719 | -1.3626352719 | 0.9637364728 |
| $w = w_{F0}$ | 0.4333333333 | 0.0 | -1.0 | 0.4333333333 |
| $w < w_{F0}$ | 0.2 | 0.2325128864 | -0.7674871136 | 0.2232512886 |
| $w = 0$ | 0.0 | 0.5 | -0.5 | 0.05 |
| $w = w_0$ | -0.0582704770 | 0.5827047698 | -0.4172952302 | 0.0 |
| $w = w_{F1}$ | -0.4333333333 | 1.0 | 0.0 | -0.3333333333 |
| $w < w_{F1}$ | -1.0 | 1.3626352719 | 0.3626352719 | -0.8637364728 |

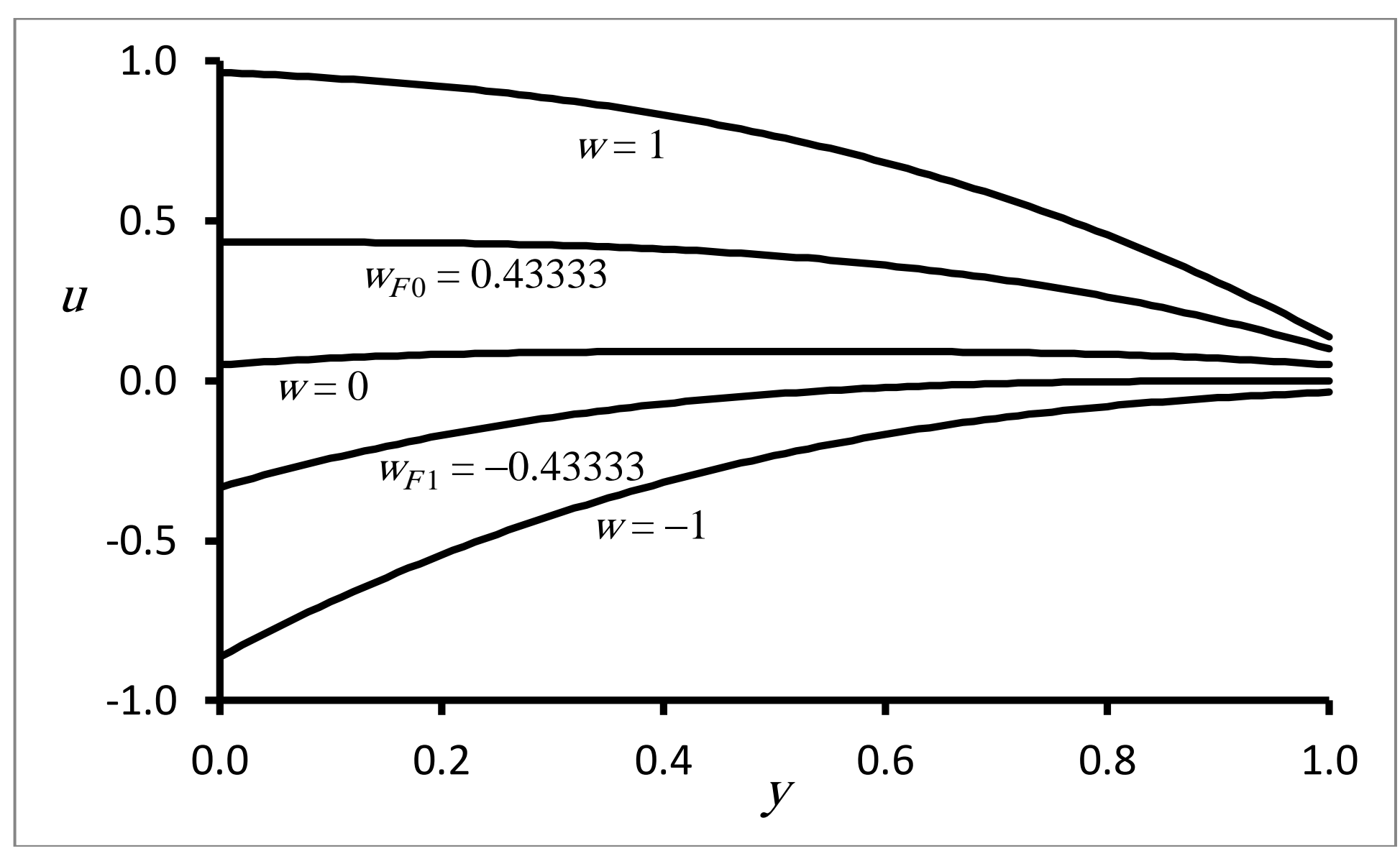

Fig. C2a: Velocity profiles for a shear-thinning fluid, $n = ½$; $\pi = 1$, $V$=0.

Table C2b: Sample results for $n = 1, \pi = 1, V = 0, \sigma_0 = \sigma_1 = 0.1$.

| Case | $w$ | $\tau_0$ | $\tau_1$ | $u_0$ |
|---|---|---|---|---|
| $w > w_{F0}$ | 1.0 | -0.3333333333 | -1.3333333333 | 0.96666666667 |
| $w = w_{F0}$ | 0.6 | 0.0 | -1.0 | 0.6 |
| $w < w_{F0}$ | 0.2 | 0.33333333333 | -0.6666666667 | 0.23333333333 |
| $w = 0$ | 0.0 | 0.5 | -0.5 | 0.05 |
| $w = w_0$ | -0.0545454545 | 0.54545454545 | -0.4545454545 | 0.0 |
| $w = w_{F1}$ | -0.6 | 1.0 | 0.0 | -0.5 |
| $w < w_{F1}$ | -1.0 | 1.3333333333 | 0.3333333333 | -0.8666666667 |

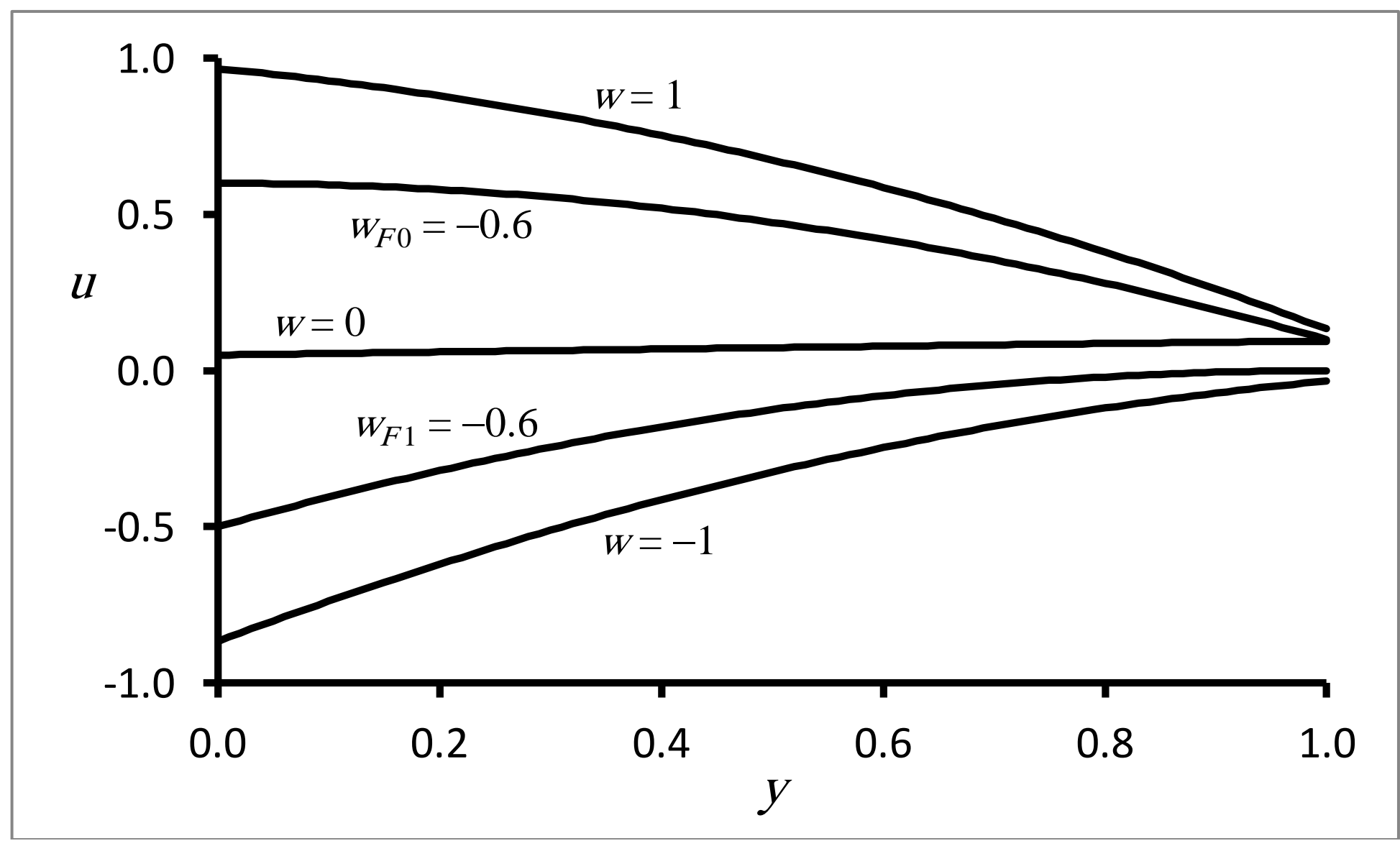

Fig. C2b: Velocity profiles for a shear-thinning fluid, $n = 1$; $\pi = 1$, $V$=0.

Table C2c: Sample results for $n = 2, \pi = 1, V = 0, \sigma_0 = \sigma_1 = 0.1$.

| Case | $w$ | $\tau_0$ | $\tau_1$ | $u_0$ |
|---|---|---|---|---|
| $w > w_{F0}$ | 1.0 | -0.2520041348 | -1.2520041348 | 0.9747995865 |
| $w = w_{F0}$ | 0.7666666667 | 0.0 | -1.0 | 0.7666666667 |
| $w < w_{F0}$ | 0.2 | 0.3758177172 | -0.6241822831 | 0.2375817717 |
| $w = 0$ | 0.0 | 0.5 | -0.5 | 0.05 |
| $w = w_0$ | -0.0533026053 | 0.5330260531 | -0.4669739469 | 0.0 |
| $w = w_{F1}$ | -0.7666666667 | 1.0 | 0.0 | -0.6666666667 |
| $w < w_{F1}$ | -1.0 | -0.8747995865 | 1.2520041348 | 0.25200413484 |

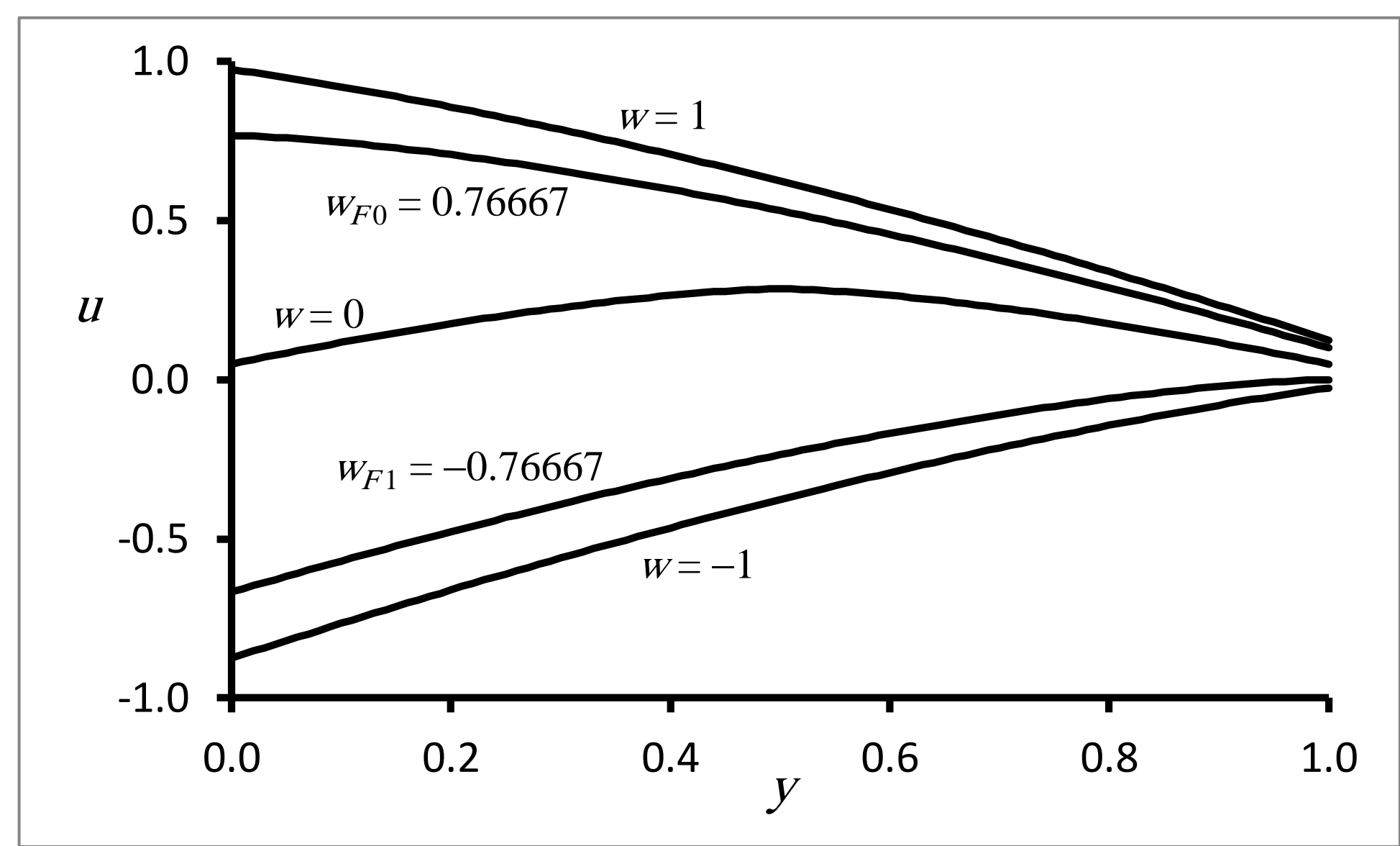

Fig. C2c: Velocity profiles for a shear-thinning fluid, $n$ =2; $\pi$ =1, $V$=0.

In Table C3, we present sample results for $w_0$ corresponding to $u_0 = 0$, when $\pi = 1$, $\sigma_0 = \sigma_1 = 0.1$.

Table C3: Sample results for $w_0$, $\pi = 1$, $\sigma_0 = \sigma_1 = 0.1$.

| $n$ | $w_0$ | $\tau_0$ | $\tau_1$ | $u_1$ | $V$ |
|---|---|---|---|---|---|
| ½ | -0.0539628194 | 0.5396281941 | -0.4185198236 | 0.04185198236 | 1 |
| 1 | -0.0447266737 | 0.4472667371 | -0.5024847844 | 0.05024847844 | |
| 2 | -0.0347534408 | 0.3475344084 | -0.5931505378 | 0.05931505378 | |
| ½ | -0.0582704770 | 0.5827047698 | -0.4172952302 | 0.04172952302 | 0 |
| 1 | -0.0545454545 | 0.54545454545 | -0.4545454545 | 0.04545454545 | |
| 2 | -0.0533026053 | 0.5330260531 | -0.4669739469 | 0.04669739469 | |